\documentclass[notitlepage,a4paper,aps,prd,onecolumn,superscriptaddress,nofootinbib,groupedaddress]{revtex4}

\usepackage{amsmath}
\usepackage{amsfonts,color}
\usepackage{amsmath}
\usepackage{amsthm}
\usepackage{pdfpages}
\usepackage{amsmath}
\usepackage{empheq}
\usepackage{amsfonts,color}
\usepackage{amssymb,float}
\usepackage{physics}
\usepackage{booktabs,multirow}
\usepackage{titlesec}

\usepackage{hyphenat}
\usepackage{ragged2e}

\titleformat{\paragraph}
{\normalfont\normalsize\bfseries}{\theparagraph}{1em}{}
\titlespacing*{\paragraph}
{0pt}{3.25ex plus 1ex minus .2ex}{1.5ex plus .2ex}

\usepackage{amsmath}
\usepackage[utf8]{inputenc}
\allowdisplaybreaks
\usepackage{color}
\usepackage{hyperref}
\hypersetup{
    colorlinks=true,
    linkcolor=blue,
    filecolor=magenta,      
   citecolor=blue
}

\usepackage{enumitem}

\usepackage{tikz}
\usetikzlibrary{shapes.geometric}
\usetikzlibrary{arrows.meta,arrows}

\begin{document}
\title{Algebraic classification of the gravitational field in Weyl-Cartan space-times}

\author{Sebastian Bahamonde}
\email{bahamonde.s.aa@m.titech.ac.jp}
\affiliation{Department of Physics, Tokyo Institute of Technology
1-12-1 Ookayama, Meguro-ku, Tokyo 152-8551, Japan.}

\author{Jorge Gigante Valcarcel}
\email{gigante.j.aa@m.titech.ac.jp, jorge.gigante.valcarcel@ut.ee}
\affiliation{Department of Physics, Tokyo Institute of Technology
1-12-1 Ookayama, Meguro-ku, Tokyo 152-8551, Japan.}
\affiliation{Laboratory of Theoretical Physics, Institute of Physics, University of Tartu, W. Ostwaldi 1, 50411 Tartu, Estonia.}

\begin{abstract}

We present a complete algebraic classification for the curvature tensor in Weyl-Cartan geometry, by applying methods of eigenvalues and principal null directions on its irreducible decomposition under the group of global Lorentz transformations, thus providing a full invariant characterisation of all the possible algebraic types of the torsion and nonmetricity field strength tensors in Weyl-Cartan space-times. As an application, we show that in the framework of Metric-Affine Gravity the field strength tensors of a dynamical torsion field cannot be doubly aligned with the principal null directions of the Riemannian Weyl tensor in scalar-flat, slowly rotating, stationary and axisymmetric space-times.

\end{abstract}

\maketitle

\section{Introduction}

Certainly, General Relativity (GR) stands out as one of the most fundamental and accurate theories in Physics, describing gravitational interaction as a geometrical property of the space-time. This success lies on a large number of theoretical and phenomenological foundations, which meet together in the celebrated Einstein's field equations~\cite{Wald:1984rg}. The significance of these equations is remarkable, since gravitational phenomena turns out to be governed by a fundamental correspondence between the space–time geometry and the energy-momentum current of matter, both described by tensor quantities on a Riemannian (Lorentzian) manifold. This fact highlights the importance of the search and the analysis of solutions of the field equations, in order to shed light on the physical implications of the theory and to test it experimentally~\cite{Stephani:2003tm,Griffiths:2009dfa}.

Nevertheless, the nonlinear behaviour of the field equations represents a significant obstacle for their resolution and for the accurate understanding of their solutions. In this sense, the role of algebraic classification has shown to be crucial to simplify the complexity of the equations and to unravel important properties of their solutions, for a large number of especially symmetric configurations (e.g. see~\cite{goldberg1962theorem} for the derivation of the Goldberg-Sachs theorem, which set the stage for the finding of the Kerr solution of GR~\cite{Kerr:1963ud}).

From a mathematical point of view, an algebraic classification of a tensor quantity can be obtained by analysing its algebraic properties and its structure as linear maps on a certain vector space by means of an eigenvalue problem, whose resolution provides a set of geometric multiplicities and the different types that compose the classification itself~\cite{petrov2016einstein}. In this regard, the most relevant classifications in four-dimensional Riemannian manifolds with Lorentzian signature are the Petrov classification of the Weyl tensor and the Segre classification of the Ricci tensor~\cite{Petrov:2000bs,bona1992intrinsic} (see~\cite{Coley:2004jv,Milson:2004jx,Ortaggio:2012jd} for further generalisations in higher dimensions). Indeed, they provide a full invariant characterisation of the gravitational field described in Riemannian geometry, in virtue of the Ricci decomposition of the Riemann curvature tensor~\cite{misner1973gravitation}. For the particular case of GR, the Einstein's field equations further imply that the energy-momentum tensor of matter must present the same class of algebraic classification as the Ricci tensor, which additionally simplifies the final form of the algebraic classification in Riemannian geometry under reasonable energy conditions~\cite{Penrose:1986ca}.

Accordingly, the set-up of algebraic classification in GR and other theories of gravity based on Riemannian geometry is clear, but the presence of different degrees of freedom in the geometry requires an extension of these results. In particular, the application of the gauge principles in the framework of post-Riemannian geometry displays the torsion and nonmetricity tensors as new features of the gravitational field, which leads to the formulation of Metric-Affine Gauge theory of gravity (MAG), as an extension of GR~\cite{Hehl:1994ue}. An action principle can then be defined from the generalised field strength tensors of this framework to introduce the dynamics of the torsion and nonmetricity fields, which provides a rich and diverse range of gravitational phenomena~\cite{adamowicz1980plane,Baekler:1981lkh,Gonner:1984rw,Bakler:1988nq,Gladchenko:1994wu,Tresguerres:1995js,tresguerres1995exact,Hehl:1999sb,Garcia:2000yi,Puetzfeld:2001hk,Chen:2009at,Lu:2016bcx,Cembranos:2016gdt,Cembranos:2017pcs,Blagojevic:2017wzf,Blagojevic:2017ssv,Obukhov:2019fti,Olmo:2019flu,Guerrero:2020azx,Bahamonde:2020fnq,Obukhov:2020hlp,Iosifidis:2020gth,Aoki:2020zqm,Iosifidis:2020upr,Bahamonde:2021akc,Bahamonde:2021qjk,delaCruzDombriz:2021nrg,Jimenez-Cano:2022arz,Bahamonde:2022meb,Bombacigno:2022lcx,Boudet:2022wmb,Battista:2022sci,Boudet:2022nub,Bahamonde:2022kwg,Yang:2022icz,Iosifidis:2022xvp,Battista:2023znv,Elizalde:2022vvc,Iosifidis:2023pvz,Ditta:2023wye,Yasir:2023qho}.

Following these lines, in this work we address this problem and present the algebraic classification for the different field strength tensors of the gravitational field in the framework of Weyl-Cartan geometry, which is characterised by the presence of curvature, torsion and nonmetricity, the latter being determined by the so-called Weyl vector. For this task, in Sec.~\ref{sec:metricaffine} we give a brief introduction of metric-affine geometry, specifying the main geometrical quantities and conventions, as well as the irreducible decomposition of the curvature tensor, which shall provide the building blocks for the classification. In Sec.~\ref{sec:classification} we focus on Weyl-Cartan space-times and present the algebraic classification for all the irreducible parts of the curvature tensor, which contain the field strength tensors of the torsion and nonmetricity fields. Then, in Sec.~\ref{sec:application} we apply these results to the search of exact black hole solutions in axial symmetry, pointing out the absence of nontrivial solutions in the torsion sector with field strength tensors that are doubly aligned to the principal null directions of the Riemannian Weyl tensor in scalar-flat stationary and axisymmetric space-times, under the slow-rotation approximation. We end with summarising conclusions in section Sec.~\ref{sec:conclusions}.

We work in natural units $c=G=1$, and we consider the metric signature $(+,-,-,-)$. On the other hand, we use a tilde accent to denote quantities defined from the general affine connection, in contrast to their unaccented counterparts constructed only from the Levi-Civita connection. For classification purposes, we denote with a diagonal arrow the traceless and pseudotraceless pieces of tensors (e.g. ${\nearrow\!\!\!\!\!\!\!\tilde{R}}\,^{\lambda}{}_{[\rho\mu\nu]}$), in order to distinguish these pieces from their counterparts containing trace and pseudotrace parts. Uppercase and lowercase Latin indices run from $0$ to $3$, referring to spinor and anholonomic indices, respectively. The same applies for Greek indices, which refer to coordinate bases.
\begin{RaggedRight}
As for the computational methods applied, we include as a supplemental material a code based on {\it Wolfram Mathematica} using the tensor algebra suite of packages {\it xAct}~\cite{martin2020xact}\end{RaggedRight}, which can be used to obtain and verify the form of all the irreducible parts of the curvature tensor from the building blocks of the decomposition, as well as the relation of these building blocks with the torsion and nonmetricity tensors described in MAG~\cite{SM}.

\section{Introduction of metric-affine geometry}\label{sec:metricaffine}

In this section, we will provide a basic introduction to metric-affine geometry and provide a full characterisation of the independent components of the curvature tensor when torsion and nonmetricity are nonvanishing.

\subsection{Main geometrical quantities and conventions}

Metric-affine geometry deals with the study of differential geometry on affinely connected metric manifolds, which are characterised by a metric tensor and an independent affine connection. Specifically, an independent affine connection displays the torsion and nonmetricity tensors
\begin{equation}
    T^{\lambda}\,_{\mu \nu}=2\tilde{\Gamma}^{\lambda}\,_{[\mu \nu]}\,, \quad Q_{\lambda \mu \nu}=\tilde{\nabla}_{\lambda}g_{\mu \nu}\,,
\end{equation}
as the antisymmetric part of the affine connection and the covariant derivative of the metric tensor, respectively. These quantities are encoded into the affine connection in terms of the so-called distortion tensor
\begin{equation}
    N^{\lambda}\,_{\mu\nu}=\frac{1}{2}\left(T^{\lambda}\,_{\mu \nu}-T_{\mu}\,^{\lambda}\,_{\nu}-T_{\nu}\,^{\lambda}\,_{\mu}\right)+\frac{1}{2}\left(Q^{\lambda}\,_{\mu \nu}-Q_{\mu}\,^{\lambda}\,_{\nu}-Q_{\nu}\,^{\lambda}\,_{\mu}\right)\,,
\end{equation}
which gives rise to a generalised curvature tensor that can be expressed as the sum of the Riemann tensor and additional post-Riemannian corrections
\begin{equation}\label{totalcurvature}
\tilde{R}^{\lambda}\,_{\rho\mu\nu}=R^{\lambda}\,_{\rho\mu\nu}+\nabla_{\mu}N^{\lambda}\,_{\rho \nu}-\nabla_{\nu}N^{\lambda}\,_{\rho \mu}+N^{\lambda}\,_{\sigma \mu}N^{\sigma}\,_{\rho \nu}-N^{\lambda}\,_{\sigma \nu}N^{\sigma}\,_{\rho \mu}\,.
\end{equation}

Thereby, in metric-affine geometry it is important to distinguish between the metric (Riemann) curvature tensor and the one defined on an affinely connected metric space-time, since they do not generally coincide in the presence of torsion and nonmetricity. Indeed, the latter also fulfils its own Bianchi identities
\begin{eqnarray}\tilde{R}^{\lambda}\,_{[\mu \nu \rho]}&=&\tilde{\nabla}_{[\mu}T^{\lambda}\,_{\rho\nu]}+T^{\sigma}\,_{[\mu\rho}\,T^{\lambda}\,_{\nu] \sigma}\,,\\\label{curvbianchi1}
\tilde{\nabla}_{[\sigma |}\tilde{R}^{\lambda}\,_{\rho | \mu \nu]}&=&T^{\omega}\,_{[\sigma \mu |}\tilde{R}^{\lambda}\,_{\rho \omega | \nu]}\,,\\\label{curvbianchi2}
\tilde{R}^{\left(\lambda\rho\right)}\,_{\mu\nu}&=&\tilde{\nabla}_{[\nu}Q_{\mu]}\,^{\lambda\rho}+\frac{1}{2}\,T^{\sigma}\,_{\mu\nu}Q_{\sigma}\,^{\lambda\rho}\,.\label{curvbianchi3}
\end{eqnarray}

In addition, there are three independent traces constructed from the contraction of two indices of the curvature tensor, namely the Ricci and co-Ricci tensors
\begin{equation}
\tilde{R}_{\mu\nu}=\tilde{R}^{\lambda}\,_{\mu \lambda \nu}\,, \quad \hat{R}_{\mu\nu}=\tilde{R}_{\mu}\,^{\lambda}\,_{\nu\lambda}\,,
\end{equation}
as well as the so-called homothetic curvature tensor
\begin{equation}
    \tilde{R}^{\lambda}\,_{\lambda\mu\nu}=\delta^{\rho}{}_{\lambda}\tilde{R}^{\lambda}\,_{\rho\mu\nu}\,.
\end{equation}

It is worthwhile to note that the trace of the Ricci and co-Ricci tensors provides a unique independent scalar curvature
\begin{equation}
    \tilde{R}=g^{\mu\nu}\tilde{R}_{\mu\nu}\,,
\end{equation}
whereas the contraction of the curvature tensor with the Levi-Civita tensor gives rise to the Holst pseudoscalar
\begin{equation}
\ast\tilde{R}=\varepsilon^{\lambda\rho\mu\nu}\tilde{R}_{\lambda\rho\mu\nu}\,.
\end{equation}

A gauge approach to gravity can then be consistently formulated by the introduction of a principal bundle with its proper local metric structure, tetrad field and connection~\cite{Hehl:1994ue}:
\begin{eqnarray}
g_{\mu \nu}&=&e^{a}\,_{\mu}\,e^{b}\,_{\nu}\,g_{a b}\,,\\
\label{anholonomic_connection}
\omega^{a}\,_{b\mu}&=&e^{a}\,_{\lambda}\,e_{b}\,^{\rho}\,\tilde{\Gamma}^{\lambda}\,_{\rho \mu}+e^{a}\,_{\lambda}\,\partial_{\mu}\,e_{b}\,^{\lambda}\,,
\end{eqnarray}
which provides a solid mathematical framework to analyse the space-time curvature, torsion and nonmetricity. In this regard, the choice of basis vectors to evaluate the main properties and the physical implications of these quantities is a matter of convenience. In particular, a null basis composed by two real null vectors $k^{\mu}$ and $l^{\mu}$, as well as by two complex conjugate null vectors $m^{\mu}$ and $\bar{m}^{\mu}$, fulfilling the pseudo-orthogonality and normalisation conditions
\begin{eqnarray}
    k^\mu l_\mu&=&-\,m^\mu \bar{m}_\mu=1\,,\label{ort1}\\
    k^\mu m_\mu&=& k^\mu \bar{m}_\mu=l^\mu m_\mu= l^\mu \bar{m}_\mu=0\,,\\
    k^\mu k_\mu&=& l^\mu l_\mu= m^\mu m_\mu= \bar{m}^\mu \bar{m}_\mu=0\,,\label{ort3}
\end{eqnarray}
has shown to be especially convenient for the study of the inherent symmetries and the congruences of null curves of the space-time~\cite{Penrose:1986ca}.

Following these lines, a gravitational action can be defined from the gauge principles, in such a way that higher-order
curvature corrections are required in the Lagrangian to endow the torsion and nonmetricity fields with dynamics~\cite{Gronwald:1995em}. In particular, the most general parity conserving quadratic Lagrangian includes $3 + 4$ irreducible parts from torsion and nonmetricity, as well as $11$ ones from curvature, which provide the dynamics and act as field strength tensors of the gravitational field in MAG.

\subsection{Irreducible decomposition of the curvature tensor}

As mentioned in the previous section, the curvature tensor encodes the field strength tensors of the gravitational field in the framework of MAG. Therefore, in order to perform an algebraic classification of the respective field strength tensors, it is essential to decompose first the curvature tensor of an affinely connected metric space-time into its irreducible pieces under the group of global Lorentz transformations~\cite{McCrea:1992wa}. In general, there exist $11$ irreducible pieces arising from this algebraic procedure, which can be grouped into the antisymmetric and symmetric components of the curvature tensor
\begin{equation}\label{irrdec}
    \tilde{R}_{\lambda\rho\mu\nu}= \tilde{W}_{\lambda\rho\mu\nu}+\tilde{Z}_{\lambda\rho\mu\nu}\,,
\end{equation}
with
\begin{equation}
\tilde{W}_{\lambda\rho\mu\nu}:=\tilde{R}_{[\lambda\rho]\mu\nu}\,,\quad \tilde{Z}_{\lambda\rho\mu\nu}:=\tilde{R}_{(\lambda\rho)\mu\nu}\,.
\end{equation}

Prior to the introduction of the irreducible decomposition of each component, it is worthwhile to define the following quantities, which shall be part of the building blocks of the respective irreducible pieces:
\begin{eqnarray}
   {\nearrow\!\!\!\!\!\!\!\tilde{R}}_{(\mu\nu)}&=&\tilde{R}_{(\mu\nu)}-\frac{1}{4}g_{\mu\nu}\tilde{R}\,, \quad {\nearrow\!\!\!\!\!\!\!\hat{R}}_{(\mu\nu)}^{(Q)}=\hat{R}_{(\mu\nu)}-\tilde{R}_{(\mu\nu)}\,,\label{first_definition}\\
\tilde{R}^{(T)}_{[\mu\nu]}&=&\tilde{R}_{[\mu \nu]}-\frac{1}{2}\tilde{R}^\lambda\,_{\lambda\mu\nu}\,, \quad \hat{R}^{(Q)}_{[\mu\nu]}=\hat{R}_{[\mu\nu]}-\tilde{R}^{(T)}_{[\mu\nu]}\,,\\
    {\nearrow\!\!\!\!\!\!\!\tilde{R}}^{(T)}_{\lambda[\rho\mu\nu]}&=&\tilde{R}_{\lambda[\rho\mu\nu]}+\frac{1}{3}\left(g_{\lambda\rho}\tilde{R}^{(T)}_{[\mu\nu]}+g_{\lambda\mu}\tilde{R}^{(T)}_{[\nu\rho]}+g_{\lambda\nu}\tilde{R}^{(T)}_{[\rho\mu]}\right)+\frac{1}{24}\left.\ast\tilde{R}\right. \varepsilon_{\lambda\rho\mu\nu}\,,\\
    {\nearrow\!\!\!\!\!\!\!\tilde{R}}_{(\lambda\rho)\mu\nu}&=&\tilde{R}_{(\lambda\rho)\mu\nu}-\frac{1}{4}g_{\lambda\rho}\tilde{R}^{\sigma}{}_{\sigma\mu\nu}+\frac{1}{6}\left(g_{\lambda\mu}\hat{R}^{(Q)}_{[\rho\nu]}+g_{\rho\mu}\hat{R}^{(Q)}_{[\lambda\nu]}-g_{\lambda \nu}\hat{R}^{(Q)}_{[\rho\mu]}-g_{\rho\nu}\hat{R}^{(Q)}_{[\lambda\mu]}-g_{\lambda\rho}\hat{R}^{(Q)}_{[\mu\nu]}\right)+\nonumber\\
&&+\,\frac{1}{8}
\left(g_{\lambda\mu}{\nearrow\!\!\!\!\!\!\!\hat{R}}^{(Q)}_{(\rho\nu)}+g_{\rho\mu}{\nearrow\!\!\!\!\!\!\!\hat{R}}^{(Q)}_{(\lambda\nu)}-g_{\lambda\nu}{\nearrow\!\!\!\!\!\!\!\hat{R}}^{(Q)}_{(\rho\mu)}-g_{\rho\nu}{\nearrow\!\!\!\!\!\!\!\hat{R}}^{(Q)}_{(\lambda\mu)}\right)\,,\\
{\nearrow\!\!\!\!\!\!\!\tilde{R}}^{(Q)}_{\lambda[\rho\mu\nu]} &=&{\nearrow\!\!\!\!\!\!\!\tilde{R}}_{(\lambda\rho)\mu\nu}+{\nearrow\!\!\!\!\!\!\!\tilde{R}}_{(\lambda\mu)\nu\rho}+{\nearrow\!\!\!\!\!\!\!\tilde{R}}_{(\lambda\nu)\rho\mu}\,.\label{last_definition}
\end{eqnarray}

We distinguish with a superscript $T$ or $Q$ those quantities, defined from the curvature tensor and its contractions, that acquire nontrivial values due to torsion or nonmetricity, respectively (see the complete list of values of the building blocks of the curvature decomposition in Appendix~\ref{sec:buildingblocks}).

\subsubsection{Antisymmetric part}

In general, the antisymmetric component of the curvature tensor can be decomposed into six irreducible pieces
\begin{eqnarray}
  \tilde{W}_{\lambda\rho\mu\nu}=\sum_{i=1}^{6}{}^{(i)}\tilde{W}_{\lambda\rho\mu\nu}\,,
\end{eqnarray}
namely
\begin{eqnarray}
{}^{(1)}\tilde{W}_{\lambda\rho\mu\nu}&=&\tilde{W}_{\lambda\rho\mu\nu}-\sum_{i=2}^{6}{}^{(i)}\tilde{W}_{\lambda\rho\mu\nu}\,,\label{first_irreducible_antisymmetric_piece}\\
{}^{(2)}\tilde{W}_{\lambda\rho\mu\nu}&=&\frac{3}{4}\Big({\nearrow\!\!\!\!\!\!\!\tilde{R}}^{(T)}_{\lambda[\rho\mu\nu]}+{\nearrow\!\!\!\!\!\!\!\tilde{R}}^{(T)}_{\nu[\lambda\rho\mu]}-{\nearrow\!\!\!\!\!\!\!\tilde{R}}^{(T)}_{\rho[\lambda\mu\nu]}-{\nearrow\!\!\!\!\!\!\!\tilde{R}}^{(T)}_{\mu[\lambda\rho\nu]}\Bigr)+\frac{1}{2}\Bigl({\nearrow\!\!\!\!\!\!\!\tilde{R}}^{(Q)}_{\mu[\lambda\rho\nu]}-{\nearrow\!\!\!\!\!\!\!\tilde{R}}^{(Q)}_{\nu[\lambda\rho\mu]}\Big)\,,\\
    {}^{(3)}\tilde{W}_{\lambda\rho\mu\nu}&=&-\,\frac{1}{24}\left.\ast\tilde{R}\right. \varepsilon_{\lambda\rho\mu\nu}\,,\\
     {}^{(4)}\tilde{W}_{\lambda\rho\mu\nu}&=&\frac{1}{4}\Big[g_{\lambda\mu}\left(2{\nearrow\!\!\!\!\!\!\!\tilde{R}}_{(\rho\nu)}+{\nearrow\!\!\!\!\!\!\!\hat{R}}^{(Q)}_{(\rho \nu) }\right)+g_{\rho\nu}\left(2{\nearrow\!\!\!\!\!\!\!\tilde{R}}_{(\lambda\mu)} +{\nearrow\!\!\!\!\!\!\!\hat{R}}^{(Q)}_{(\lambda\mu)}\right)-g_{\lambda\nu}\left(2 {\nearrow\!\!\!\!\!\!\!\tilde{R}}_{(\rho\mu)}+{\nearrow\!\!\!\!\!\!\!\hat{R}}^{(Q)}_{(\rho\mu)}\right)-g_{\rho\mu}\left(2{\nearrow\!\!\!\!\!\!\!\tilde{R}}_{(\lambda\nu)}+{\nearrow\!\!\!\!\!\!\!\hat{R}}^{(Q)}_{(\lambda\nu)}\right)\Big]\,,\nonumber \\ \\
{}^{(5)}\tilde{W}_{\lambda\rho\mu\nu}&=&\frac{1}{4}\Big[g_{\lambda\mu}\Bigl(2\tilde{R}^{(T)}_{[\rho\nu]}+\hat{R}^{(Q)}_{[\rho\nu]}\Bigr)+g_{\rho\nu}\Bigl(2\tilde{R}^{(T)}_{[\lambda\mu]}+\hat{R}^{(Q)}_{[\lambda\mu]}\Bigr)-g_{\lambda\nu}\Bigl(2\tilde{R}^{(T)}_{[\rho\mu]}+\hat{R}^{(Q)}_{[\rho\mu]}\Bigr)-g_{\rho\mu}\Bigl(2\tilde{R}^{(T)}_{[\lambda\nu]}+\hat{R}^{(Q)}_{[\lambda\nu]}\Bigr)\nonumber\\
 &&+\,\tilde{R}^{\sigma}{}_{\sigma\lambda[\mu}g_{\nu]\rho}-\tilde{R}^{\sigma}{}_{\sigma\rho[\mu}g_{\nu]\lambda}\Big]\,,\\
   {}^{(6)}\tilde{W}_{\lambda\rho\mu\nu}&=&\frac{1}{6}\,\tilde{R}\,g_{\lambda[\mu}g_{\nu]\rho}\,,\label{last_irreducible_antisymmetric_piece}
\end{eqnarray}

As is shown, the presence of torsion and nonmetricity induces post-Riemannian corrections in the three irreducible pieces ${}^{(1)}\tilde{W}_{\alpha\beta\mu\nu}$, ${}^{(4)}\tilde{W}_{\alpha\beta\mu\nu}$ and ${}^{(6)}\tilde{W}_{\alpha\beta\mu\nu}$, which in fact include the Weyl, traceless Ricci and scalar parts of the curvature tensor of Riemannian geometry. Indeed, the tensor ${}^{(1)}\tilde{W}_{\lambda\rho\mu\nu}$ presents the same algebraic properties as the Riemannian Weyl tensor:
\begin{align}
    {}^{(1)}\tilde{W}_{\lambda\rho\mu\nu}&=-\,{}^{(1)}\tilde{W}_{\rho\lambda\mu\nu}=-\,{}^{(1)}\tilde{W}_{\lambda\rho\nu\mu}\,,\label{Piece1_property1}\\
    {}^{(1)}\tilde{W}_{\lambda[\rho\mu\nu]}&={}^{(1)}\tilde{W}^{\lambda}{}_{\mu\lambda\nu}=0\,,\label{Piece1_property2}
\end{align}
and can be equivalently expressed as follows:
\begin{eqnarray}
    {}^{(1)}\tilde{W}_{\lambda\rho\mu\nu}&=&\frac{1}{2}\bigl(\tilde{R}_{[\lambda\rho]\mu\nu}+\tilde{R}_{[\mu\nu]\lambda\rho}\bigr)+\frac{1}{4}\big[g_{\mu\rho}\bigl(\tilde{R}_{(\lambda\nu)}+\hat{R}_{(\lambda\nu)}\bigr)-g_{\mu\lambda}\bigl(\tilde{R}_{(\rho\nu)}+\hat{R}_{(\rho\nu)}\bigr)+g_{\nu\lambda}\bigl(\tilde{R}_{(\rho\mu)}+\hat{R}_{(\rho\mu)}\bigr)\nonumber\\
    &&-\,g_{\nu\rho}\bigl(\tilde{R}_{(\lambda\mu)}+\hat{R}_{(\lambda\mu)}\bigr)\big]+\frac{1}{3}\tilde{R}g_{\lambda[\mu}g_{\nu]\rho}+\frac{1}{24}\left.\ast\tilde{R}\right.\varepsilon_{\lambda\rho\mu\nu}\,.
\end{eqnarray}
Clearly, when torsion and nonmetricity are vanishing, the tensor ${}^{(1)}\tilde{W}_{\lambda\rho\mu\nu}$ reduces to the Riemannian Weyl tensor.

In addition, three purely post-Riemannian pieces ${}^{(2)}\tilde{W}_{\alpha\beta\mu\nu}$, ${}^{(3)}\tilde{W}_{\alpha\beta\mu\nu}$ and ${}^{(5)}\tilde{W}_{\alpha\beta\mu\nu}$ arise, thus settling the irreducible decomposition of the antisymmetric component of the curvature tensor.

\subsubsection{Symmetric part}

The symmetric component of the curvature tensor only takes nontrivial values in the presence of nonmetricity and can be decomposed into five irreducible pieces:
\begin{eqnarray}
   \tilde{Z}_{\lambda\rho\mu\nu}=\sum_{i=1}^{5}{}^{(i)}\tilde{Z}_{\lambda\rho\mu\nu}\,,
\end{eqnarray}
where
\begin{eqnarray}
 {}^{(1)}\tilde{Z}_{\lambda\rho\mu\nu}&=& \tilde{Z}_{\lambda\rho\mu\nu}-\sum_{i=2}^{5}{}^{(i)}\tilde{Z}_{\lambda\rho\mu\nu}\,,\label{first_irreducible_symmetric_piece}\\
       {}^{(2)}\tilde{Z}_{\lambda\rho\mu\nu}&=&\frac{1}{4}\Big({\nearrow\!\!\!\!\!\!\!\tilde{R}}^{(Q)}_{\lambda[\rho\mu\nu]}+{\nearrow\!\!\!\!\!\!\!\tilde{R}}^{(Q)}_{\rho[\lambda\mu\nu]}\Big)\,,\\
       {}^{(3)}\tilde{Z}_{\lambda\rho\mu\nu}&=&\frac{1}{6}\Big(g_{\lambda\nu}\hat{R}^{(Q)}_{[\rho\mu]}+g_{\rho\nu}\hat{R}^{(Q)}_{[\lambda\mu]}-g_{\lambda\mu}\hat{R}^{(Q)}_{[\rho\nu]}-g_{\rho\mu}\hat{R}^{(Q)}_{[\lambda\nu]}+g_{\lambda\rho}\hat{R}^{(Q)}_{[\mu\nu]}\Big)\,,\\
       {}^{(4)}\tilde{Z}_{\lambda\rho\mu\nu}&=&\frac{1}{4}g_{\lambda\rho}\tilde{R}^{\sigma}{}_{\sigma\mu\nu}\,,\\
       {}^{(5)}\tilde{Z}_{\lambda\rho\mu\nu}&=&\frac{1}{8}\Big(g_{\lambda\nu}{\nearrow\!\!\!\!\!\!\!\hat{R}}^{(Q)}_{(\rho\mu)}+g_{\rho\nu} {\nearrow\!\!\!\!\!\!\!\hat{R}}^{(Q)}_{(\lambda\mu)}-g_{\lambda\mu} {\nearrow\!\!\!\!\!\!\!\hat{R}}^{(Q)}_{(\rho\nu)}-g_{\rho\mu} {\nearrow\!\!\!\!\!\!\!\hat{R}}^{(Q)}_{(\lambda\nu)}\Big)\,.\label{last_irreducible_symmetric_piece}
\end{eqnarray}

The first piece ${}^{(1)}\tilde{Z}_{\lambda\rho\mu\nu}$ fulfils the algebraic properties
\begin{eqnarray}
    {}^{(1)}\tilde{Z}_{\lambda[\rho\mu\nu]}=0\,,\label{conditionZ}\quad
    {}^{(1)}\tilde{Z}^{\lambda}{}_{\lambda\mu\nu}=0\,,\quad
    {}^{(1)}\tilde{Z}^{\lambda}{}_{\mu\lambda\nu}=0\,,
\end{eqnarray}
and can also be written as
\begin{align}
{}^{(1)}\tilde{Z}_{\lambda\rho\mu\nu}&= \frac{1}{4}\Big(2\tilde{R}_{(\lambda\rho)\mu\nu}+\tilde{R}_{(\lambda\nu)\mu\rho}-\tilde{R}_{(\lambda\mu)\nu\rho}+\tilde{R}_{(\rho\mu)\lambda\nu}-\tilde{R}_{(\rho\nu)\lambda\mu}\Big)-\frac{1}{12}g_{\lambda\rho}(\tilde{R}^\alpha{}_{\alpha\mu\nu}+\tilde{R}_{[\mu\nu]}-\hat{R}_{[\mu\nu]})\nonumber\\
&-\frac{1}{24}g_{\lambda\mu}\Big(\tilde{R}^\alpha{}_{\alpha\rho\nu}+\tilde{R}_{[\rho\nu]}-\hat{R}_{[\rho\nu]}+3(\tilde{R}_{(\rho\nu)}-\hat{R}_{(\rho\nu)})\Big)+\frac{1}{24}g_{\lambda\nu}\Big(\tilde{R}^\alpha{}_{\alpha\rho\mu}+\tilde{R}_{[\rho\mu]}-\hat{R}_{[\rho\mu]}+3(\tilde{R}_{(\rho\mu)}-\hat{R}_{(\rho\mu)})\Big)\nonumber\\
&-\frac{1}{24}g_{\rho\mu}\Big(\tilde{R}^\alpha{}_{\alpha\lambda\nu}+\tilde{R}_{[\lambda\nu]}-\hat{R}_{[\lambda\nu]}+3(\tilde{R}_{(\lambda\nu)}-\hat{R}_{(\lambda\nu)})\Big)+\frac{1}{24}g_{\rho\nu}\Big(\tilde{R}^\alpha{}_{\alpha\lambda\mu}+\tilde{R}_{[\lambda\mu]}-\hat{R}_{[\lambda\mu]}+3(\tilde{R}_{(\lambda\mu)}-\hat{R}_{(\lambda\mu)})\Big)\,.
\end{align}
Likewise, the second piece ${}^{(2)}\tilde{Z}_{\lambda\rho\mu\nu}$ satisfies
\begin{eqnarray}
    {}^{(2)}\tilde{Z}_{(\lambda\rho\mu)\nu}=0\,,\label{conditionZ2}\quad
    {}^{(2)}\tilde{Z}^{\lambda}{}_{\lambda\mu\nu}=0\,,\quad
    {}^{(2)}\tilde{Z}^{\lambda}{}_{\mu\lambda\nu}=0\,.
\end{eqnarray}

Since the symmetric component of the curvature tensor is nonvanishing only in the presence of nonmetricity, it is worthwhile to take into account that the nonmetricity tensor can also be separated as the sum of two trace and traceless parts
\begin{equation}
Q_{\lambda\mu\nu}=\frac{1}{4}g_{\mu\nu}Q_{\lambda\rho}\,^{\rho}+{\nearrow\!\!\!\!\!\!\!Q}_{\lambda\mu\nu}\,,
\end{equation}
the former being related to the Weyl vector
\begin{equation}\label{Weyl_vector}
    W_{\mu}=\frac{1}{4}\,Q_{\mu\nu}\,^{\nu}\,.
\end{equation}
Then, it turns out that the homothetic part of the curvature tensor is fully determined by the Weyl vector
\begin{equation}
    \tilde{R}^{\lambda}\,_{\lambda\mu\nu}=4\nabla_{[\nu}W_{\mu]}\,,
\end{equation}
whereas the rest of quantities in the irreducible decomposition of the symmetric component of the curvature tensor are nonzero only in the presence of a traceless part of the nonmetricity tensor. In particular, Weyl-Cartan space-times are characterised by a nontrivial piece ${}^{(4)}\tilde{Z}_{\lambda\rho\mu\nu}$, being the rest of irreducible pieces of this component equal to zero.

\subsubsection{The building blocks of the irreducible decomposition}

As shown in the previous subsections, the curvature tensor can be decomposed into $11$ irreducible pieces, which provide the dynamics of the gravitational field in MAG. Accordingly, the algebraic classification in metric-affine geometry requires an analysis on the algebraic properties of each irreducible piece. For this task, it is essential to identify the building blocks that provide the decomposition and their number of independent components.

According to our definitions~\eqref{first_definition}-\eqref{last_definition}, it is straightforward to note that the building blocks of the irreducible decomposition~\eqref{first_irreducible_antisymmetric_piece}-\eqref{last_irreducible_antisymmetric_piece} and~\eqref{first_irreducible_symmetric_piece}-\eqref{last_irreducible_symmetric_piece} are the following $11$ quantities:
\begin{equation}\label{list_of_building_blocks}\Big\{^{(1)}\tilde{Z}_{\lambda\rho\mu\nu},^{(1)}\tilde{W}_{\lambda\rho\mu\nu},{\nearrow\!\!\!\!\!\!\!\tilde{R}}^{(T)}_{\lambda[\rho\mu\nu]},{\nearrow\!\!\!\!\!\!\!\tilde{R}}^{(Q)}_{\lambda[\rho\mu\nu]},{\nearrow\!\!\!\!\!\!\!\tilde{R}}_{(\mu\nu)},{\nearrow\!\!\!\!\!\!\!\hat{R}}_{(\mu\nu)}^{(Q)},\tilde{R}^{(T)}_{[\mu\nu]},\hat{R}^{(Q)}_{[\mu\nu]},\tilde{R}^\lambda\,_{\lambda\mu\nu},\tilde{R},\ast\tilde{R}\Big\}\,.\end{equation}

Therefore, as shown in Table~\ref{tab:buildingblocks}, the number of independent components of the curvature tensor $\tilde{R}_{\lambda\rho\mu\nu}$ coincides with the total number of independent components of the aforementioned building blocks, which is constrained by their algebraic properties. Specifically, for a four-dimensional affinely connected metric space-time the generalised Weyl tensor $^{(1)}\tilde{W}_{\lambda\rho\mu\nu}$ satisfies the same algebraic properties as the Riemannian Weyl tensor, which means that it contributes with $10$ independent components in the irreducible decomposition of the curvature tensor. The same applies to the symmetric traceless Ricci tensor ${\nearrow\!\!\!\!\!\!\!\tilde{R}}_{(\mu\nu)}$, which contributes then with $9$ independent components, whereas the fact that the tensor ${\nearrow\!\!\!\!\!\!\!\hat{R}}_{(\mu\nu)}^{(Q)}$ is also symmetric and traceless equally means $9$ independent components for this tensor. Furthermore, the antisymmetrised pieces ${\nearrow\!\!\!\!\!\!\!\tilde{R}}^{(T)}_{\lambda[\rho\mu\nu]}$ and ${\nearrow\!\!\!\!\!\!\!\tilde{R}}^{(Q)}_{\lambda[\rho\mu\nu]}$ also have $9$ independent components, in virtue of their traceless and pseudotraceless properties. On the other hand, the antisymmetric Ricci and co-Ricci-like tensors $\tilde{R}^{(T)}_{[\mu\nu]}$ and $\hat{R}^{(Q)}_{[\mu\nu]}$, as well as the homothetic curvature tensor $\tilde{R}^\lambda\,_{\lambda\mu\nu}$, provide $6$ independent components for each piece, whereas the Ricci scalar $\tilde{R}$ and the Holst pseudoscalar $\ast\tilde{R}$ represent $1$ degree of freedom, respectively. Finally, the fact that the symmetric component of the curvature tensor includes $60$ degrees of freedom means that the piece $^{(1)}\tilde{Z}_{\lambda\rho\mu\nu}$ represents the remaining $30$ degrees of freedom in the decomposition.

\begin{table*}
\begin{center}
    \begin{tabular}{| c | c |}
    \hline
    Building block & Number of independent components \\ \hline
    
    $^{(1)}\tilde{Z}_{\lambda\rho\mu\nu}$ & 30  \\ \hline
    
    $^{(1)}\tilde{W}_{\lambda\rho\mu\nu}$ & 10  \\ \hline
    ${\nearrow\!\!\!\!\!\!\!\tilde{R}}^{(T)}_{\lambda[\rho\mu\nu]}$ & 9  \\ \hline
    ${\nearrow\!\!\!\!\!\!\!\tilde{R}}^{(Q)}_{\lambda[\rho\mu\nu]}$ & 9  \\ \hline
    
    ${\nearrow\!\!\!\!\!\!\!\tilde{R}}_{(\mu\nu)}$ & 9  \\ \hline
    
    ${\nearrow\!\!\!\!\!\!\!\hat{R}}_{(\mu\nu)}^{(Q)}$ & 9  \\ \hline
    
    $\tilde{R}^{(T)}_{[\mu\nu]}$ & 6  \\ \hline
    
    $\hat{R}^{(Q)}_{[\mu\nu]}$ & 6  \\ \hline
    
    $\tilde{R}^\lambda\,_{\lambda\mu\nu}$ & 6  \\ \hline

    $\tilde{R}$ & 1  \\ \hline
    
    $\ast\tilde{R}$ & 1  \\ \hline
    \end{tabular}
\end{center}
\caption{Number of independent components of the building blocks.}
\label{tab:buildingblocks}
\end{table*}

Likewise, it is worthwhile to emphasise that the presence of torsion and nonmetricity provides corrections to the Riemannian Weyl, Ricci and scalar parts of the curvature tensor (i.e. to the irreducible pieces $^{(1)}W_{\lambda\rho\mu\nu}$, $^{(4)}W_{\lambda\rho\mu\nu}$ and $^{(6)}W_{\lambda\rho\mu\nu}$ of the Ricci decomposition of Riemannian geometry), which actually preserve their algebraic properties. Accordingly, in terms of algebraic classification, the corresponding algebraic types of the Riemannian and post-Riemannian versions of these tensors can generally differ for a given configuration, but all of them must belong to the same class of classification. In addition, the torsion and nonmetricity tensors themselves generate the remaining $8$ irreducible pieces of the decomposition, which are therefore purely post-Riemannian quantities. Furthermore, the set of irreducible pieces $\{^{(i)}\tilde{Z}_{\lambda\rho\mu\nu}\}_{i=1}^{5}$ arises only in the presence of nonmetricity; in particular, the piece $^{(4)}\tilde{Z}_{\lambda\rho\mu\nu}$ is connected to the Weyl vector and the rest of pieces of this set to the traceless part of the nonmetricity tensor. See Appendix~\ref{sec:buildingblocks} for the expressions of the building blocks in terms of the torsion and nonmetricity tensors.

\section{Algebraic classification in Weyl-Cartan geometry}\label{sec:classification}

For a Weyl-Cartan space-time, the traceless part of the nonmetricity tensor vanishes, which means that nonmetricity is fully determined by the Weyl vector
\begin{equation}
Q_{\lambda\mu\nu}=g_{\mu\nu}W_{\lambda}\,.
\end{equation}

This restriction represents a strong but still meaningful simplification, since it vanishes all the irreducible pieces of the symmetric component of the curvature tensor, except the piece $^{(4)}\tilde{Z}_{\lambda\rho\mu\nu}$ associated with the homothetic part, which indeed in MAG acts as the field strength tensor of the Weyl vector.

The building blocks of the irreducible decomposition of the curvature tensor on a Weyl-Cartan space-time are then described by the following $7$ quantities:
\begin{equation}\label{list_of_building_blocksWC}\Big\{^{(1)}\tilde{W}_{\lambda\rho\mu\nu},{\nearrow\!\!\!\!\!\!\!\tilde{R}}^{(T)}_{\lambda[\rho\mu\nu]},{\nearrow\!\!\!\!\!\!\!\tilde{R}}_{(\mu\nu)},\tilde{R}^{(T)}_{[\mu\nu]},\tilde{R}^\lambda\,_{\lambda\mu\nu},\tilde{R},\ast\tilde{R}\Big\}\,.\end{equation}
Therefore, the problem of algebraic classification in Weyl-Cartan geometry requires to fully classify the aforementioned building blocks.

\subsection{Algebraic classification of $^{(1)}\tilde{W}_{\lambda\rho\mu\nu}$}

Considering the algebraic properties~\eqref{Piece1_property1} and~\eqref{Piece1_property2} for the tensor $^{(1)}\tilde{W}_{\lambda\rho\mu\nu}$, it is straightforward to set the following $26$ components:
\begin{align}
      &^{(1)}\tilde{W}_{\hat{0}\hat{1}\hat{0}\hat{1}}=-\,^{(1)}\tilde{W}_{\hat{2}\hat{3}\hat{2}\hat{3}}\,; \quad ^{(1)}\tilde{W}_{\hat{0}\hat{1}\hat{0}\hat{2}}={}^{(1)}\tilde{W}_{\hat{2}\hat{3}\hat{1}\hat{3}}\,; \\
      &^{(1)}\tilde{W}_{\hat{0}\hat{1}\hat{0}\hat{3}}=-\,^{(1)}\tilde{W}_{\hat{2}\hat{3}\hat{1}\hat{2}}\,; \quad ^{(1)}\tilde{W}_{\hat{0}\hat{1}\hat{1}\hat{2}}={}^{(1)}\tilde{W}_{\hat{2}\hat{3}\hat{0}\hat{3}}\,; \\
      &^{(1)}\tilde{W}_{\hat{0}\hat{1}\hat{1}\hat{3}}=-\,^{(1)}\tilde{W}_{\hat{2}\hat{3}\hat{0}\hat{2}}\,; \quad ^{(1)}\tilde{W}_{\hat{0}\hat{1}\hat{2}\hat{3}}={}^{(1)}\tilde{W}_{\hat{2}\hat{3}\hat{0}\hat{1}}\,; \\
      &^{(1)}\tilde{W}_{\hat{0}\hat{2}\hat{0}\hat{1}}={}^{(1)}\tilde{W}_{\hat{2}\hat{3}\hat{1}\hat{3}}\,; \quad ^{(1)}\tilde{W}_{\hat{0}\hat{2}\hat{0}\hat{2}}=-\,^{(1)}\tilde{W}_{\hat{1}\hat{3}\hat{1}\hat{3}}\,; \\
      &^{(1)}\tilde{W}_{\hat{0}\hat{2}\hat{0}\hat{3}}={}^{(1)}\tilde{W}_{\hat{1}\hat{3}\hat{1}\hat{2}}\,; \quad ^{(1)}\tilde{W}_{\hat{0}\hat{2}\hat{1}\hat{2}}=-\,^{(1)}\tilde{W}_{\hat{1}\hat{3}\hat{0}\hat{3}}\,; \\
      &^{(1)}\tilde{W}_{\hat{0}\hat{2}\hat{1}\hat{3}}={}^{(1)}\tilde{W}_{\hat{1}\hat{3}\hat{0}\hat{2}}\,; \quad ^{(1)}\tilde{W}_{\hat{0}\hat{2}\hat{2}\hat{3}}={}^{(1)}\tilde{W}_{\hat{2}\hat{3}\hat{0}\hat{2}}\,; \\
      &^{(1)}\tilde{W}_{\hat{0}\hat{3}\hat{0}\hat{1}}=-\,^{(1)}\tilde{W}_{\hat{2}\hat{3}\hat{1}\hat{2}}\,; \quad ^{(1)}\tilde{W}_{\hat{0}\hat{3}\hat{0}\hat{2}}={}^{(1)}\tilde{W}_{\hat{1}\hat{3}\hat{1}\hat{2}}\,; \\
      &^{(1)}\tilde{W}_{\hat{0}\hat{3}\hat{0}\hat{3}}={}^{(1)}\tilde{W}_{\hat{1}\hat{3}\hat{1}\hat{3}}+{}^{(1)}\tilde{W}_{\hat{2}\hat{3}\hat{2}\hat{3}}\,; \quad ^{(1)}\tilde{W}_{\hat{0}\hat{3}\hat{1}\hat{2}}={}^{(1)}\tilde{W}_{\hat{1}\hat{3}\hat{0}\hat{2}}-{}^{(1)}\tilde{W}_{\hat{2}\hat{3}\hat{0}\hat{1}}\,; \\
      &^{(1)}\tilde{W}_{\hat{0}\hat{3}\hat{1}\hat{3}}={}^{(1)}\tilde{W}_{\hat{1}\hat{3}\hat{0}\hat{3}}\,; \quad ^{(1)}\tilde{W}_{\hat{0}\hat{3}\hat{2}\hat{3}}={}^{(1)}\tilde{W}_{\hat{2}\hat{3}\hat{0}\hat{3}}\,; \\
      &^{(1)}\tilde{W}_{\hat{1}\hat{2}\hat{0}\hat{1}}={}^{(1)}\tilde{W}_{\hat{2}\hat{3}\hat{0}\hat{3}}\,; \quad ^{(1)}\tilde{W}_{\hat{1}\hat{2}\hat{0}\hat{2}}=-\,^{(1)}\tilde{W}_{\hat{1}\hat{3}\hat{0}\hat{3}}\,; \\
      &^{(1)}\tilde{W}_{\hat{1}\hat{2}\hat{0}\hat{3}}={}^{(1)}\tilde{W}_{\hat{1}\hat{3}\hat{0}\hat{2}}-{}^{(1)}\tilde{W}_{\hat{2}\hat{3}\hat{0}\hat{1}}\,; \quad ^{(1)}\tilde{W}_{\hat{1}\hat{2}\hat{1}\hat{2}}=-{}\left(^{(1)}\tilde{W}_{\hat{1}\hat{3}\hat{1}\hat{3}}+{}^{(1)}\tilde{W}_{\hat{2}\hat{3}\hat{2}\hat{3}}\right)\,; \\
      &^{(1)}\tilde{W}_{\hat{1}\hat{2}\hat{1}\hat{3}}={}^{(1)}\tilde{W}_{\hat{1}\hat{3}\hat{1}\hat{2}}\,; \quad ^{(1)}\tilde{W}_{\hat{1}\hat{2}\hat{2}\hat{3}}={}^{(1)}\tilde{W}_{\hat{2}\hat{3}\hat{1}\hat{2}}\,; \\
      &^{(1)}\tilde{W}_{\hat{1}\hat{3}\hat{0}\hat{1}}=-\,^{(1)}\tilde{W}_{\hat{2}\hat{3}\hat{0}\hat{2}}\,; \quad ^{(1)}\tilde{W}_{\hat{1}\hat{3}\hat{2}\hat{3}}={}^{(1)}\tilde{W}_{\hat{2}\hat{3}\hat{1}\hat{3}}\,.
\end{align}

Thereby, for the algebraic classification of the generalised Weyl tensor in the presence of torsion and nonmetricity, we distribute its $10$ independent components in terms of the quantity
\begin{equation}
    ^{(1)}\tilde{W}^{+}_{\hat{0}a\hat{0}b}={}^{(1)}\tilde{W}_{\hat{0}a\hat{0}b}+i\ast^{(1)}\tilde{W}_{\hat{0}a\hat{0}b}\,,
\end{equation}
where
\begin{eqnarray}
^{(1)}\tilde{W}_{\hat{0}a\hat{0}b} &=& \left(
\begin{array}{ccc}
-\,^{(1)}\tilde{W}_{\hat{2}\hat{3}\hat{2}\hat{3}} & ^{(1)}\tilde{W}_{\hat{2}\hat{3}\hat{1}\hat{3}} & -\,^{(1)}\tilde{W}_{\hat{2}\hat{3}\hat{1}\hat{2}} \\
^{(1)}\tilde{W}_{\hat{2}\hat{3}\hat{1}\hat{3}} & -\,^{(1)}\tilde{W}_{\hat{1}\hat{3}\hat{1}\hat{3}} & ^{(1)}\tilde{W}_{\hat{1}\hat{3}\hat{1}\hat{2}} \\
-\,^{(1)}\tilde{W}_{\hat{2}\hat{3}\hat{1}\hat{2}} & ^{(1)}\tilde{W}_{\hat{1}\hat{3}\hat{1}\hat{2}} & ^{(1)}\tilde{W}_{\hat{1}\hat{3}\hat{1}\hat{3}}+{}^{(1)}\tilde{W}_{\hat{2}\hat{3}\hat{2}\hat{3}}
\end{array} \right)\,,
\end{eqnarray}
\begin{eqnarray}
\ast^{(1)}\tilde{W}_{\hat{0}a\hat{0}b} &=& \left(
\begin{array}{ccc}
^{(1)}\tilde{W}_{\hat{2}\hat{3}\hat{0}\hat{1}} & ^{(1)}\tilde{W}_{\hat{2}\hat{3}\hat{0}\hat{2}} & ^{(1)}\tilde{W}_{\hat{2}\hat{3}\hat{0}\hat{3}} \\
^{(1)}\tilde{W}_{\hat{2}\hat{3}\hat{0}\hat{2}} & -\,^{(1)}\tilde{W}_{\hat{1}\hat{3}\hat{0}\hat{2}} & -\,^{(1)}\tilde{W}_{\hat{1}\hat{3}\hat{0}\hat{3}} \\
^{(1)}\tilde{W}_{\hat{2}\hat{3}\hat{0}\hat{3}} & -\,^{(1)}\tilde{W}_{\hat{1}\hat{3}\hat{0}\hat{3}} & ^{(1)}\tilde{W}_{\hat{1}\hat{3}\hat{0}\hat{2}}-{}^{(1)}\tilde{W}_{\hat{2}\hat{3}\hat{0}\hat{1}}
\end{array} \right)\,,
\end{eqnarray}
and impose the eigenvalue equation for an eigenvector $v_{a}$:
\begin{equation}\label{eigenW1}
    ^{(1)}\tilde{W}^{+}_{\hat{0}a\hat{0}b}v^{b}=\lambda v_{a}\,.
\end{equation}

The resolution of this eigenvalue problem provides then the possible algebraic types of the Weyl tensor encoded in the complex matrix $^{(1)}\tilde{W}^{+}_{\hat{0}a\hat{0}b}$, whose characteristic polynomial can be written in terms of the quadratic and cubic invariants
\begin{equation}
    I={}^{(1)}\tilde{W}^{+}_{a b c d}{}^{(1)}\tilde{W}^{a b c d}\,, \quad J={}^{(1)}\tilde{W}^{+}_{a b}{}^{c d}{}^{(1)}\tilde{W}_{c d}{}^{e f}{}^{(1)}\tilde{W}_{e f}{}^{a b}\,,
\end{equation}
as follows:
\begin{equation}
    p(\lambda)=\lambda^{3}-\frac{I}{16}\lambda-\frac{J}{48}\,.
\end{equation}

The solution to the characteristic equation $p(\lambda)=0$ provides then in general $3$ distinct complex eigenvalues
\begin{align}
    \lambda_{1}&=\frac{1}{12}\Bigl(18J+3\sqrt{36J^2-3I^3}\Bigr)^{1/3}+\frac{I}{4\Bigl(18J+3\sqrt{36J^2-3I^3}\Bigr)^{1/3}}\,,\\
    \lambda_{2}&=-\,\frac{1}{24}\Bigl(18J+3\sqrt{36J^2-3I^3}\Bigr)^{1/3}-\frac{I}{8\Bigl(18J+3\sqrt{36J^2-3I^3}\Bigr)^{1/3}}\nonumber\\
    &+i\,\frac{\sqrt{3}}{2}\left[\frac{1}{12}\Bigl(18J+3\sqrt{36J^2-3I^3}\Bigr)^{1/3}-\frac{I}{4\Bigl(18J+3\sqrt{36J^2-3I^3}\Bigr)^{1/3}}\right]\,,\\\lambda_{3}&=-\,\frac{1}{24}\Bigl(18J+3\sqrt{36J^2-3I^3}\Bigr)^{1/3}-\frac{I}{8\Bigl(18J+3\sqrt{36J^2-3I^3}\Bigr)^{1/3}}\nonumber\\
    &-i\,\frac{\sqrt{3}}{2}\left[\frac{1}{12}\Bigl(18J+3\sqrt{36J^2-3I^3}\Bigr)^{1/3}-\frac{I}{4\Bigl(18J+3\sqrt{36J^2-3I^3}\Bigr)^{1/3}}\right]\,,
\end{align}
whose geometric multiplicities give rise to the different algebraic types in the classification, in such a way that the particular cases fulfilling the condition $I^{3} = 12 J^{2}$ are called algebraically special. Indeed, each special type satisfies a constraint associated with those null vectors $\{k_{\mu},l_{\mu},m_{\mu},\bar{m}_{\mu}\}$ aligned with the tensor $^{(1)}\tilde{W}_{\lambda \rho \mu \nu}$ or principal null directions
\begin{equation}
    k_{[\sigma}{}^{(1)}\tilde{W}_{\lambda] \rho \mu [\nu}k_{\omega]}k^{\rho}k^{\mu}=l_{[\sigma}{}^{(1)}\tilde{W}_{\lambda] \rho \mu [\nu}l_{\omega]}l^{\rho}l^{\mu}=m_{[\sigma}{}^{(1)}\tilde{W}_{\lambda] \rho \mu [\nu}m_{\omega]}m^{\rho}m^{\mu}=\bar{m}_{[\sigma}{}^{(1)}\tilde{W}_{\lambda] \rho \mu [\nu}\bar{m}_{\omega]}\bar{m}^{\rho}\bar{m}^{\mu}=0\,.
\end{equation}
Specifically, the $10$ indepedent components of the tensor $^{(1)}\tilde{W}_{\lambda\rho\mu\nu}$ can also be described by $5$ complex quantities $\{\Sigma_{i}\}_{i=1}^{5}$, according to the following identity involving the null vectors:
\begin{eqnarray}
\label{W1id}
    ^{(1)}\tilde{W}_{\lambda\rho\mu\nu}&=&-\,\frac{1}{2}\left(\Sigma_2+\bar{\Sigma}_2\right)\left(\{l_\lambda k_{\rho}l_{\mu}k_\nu\}+\{m_\lambda \bar{m}_{\rho}m_{\mu}\bar{m}_\nu\}\right)+\left(\Sigma_2-\bar{\Sigma}_2\right)\{l_\lambda k_{\rho}m_{\mu}\bar{m}_\nu\}\nonumber\\
    &&-\,\frac{1}{2}\left(\bar{\Sigma}_0\{k_{\lambda}m_{\rho}k_{\mu}m_\nu\}+\Sigma_0\{k_\lambda \bar{m}_{\rho}k_{\mu}\bar{m}_\nu\}\right)-\frac{1}{2}\left(\Sigma_4\{l_\lambda m_{\rho}l_{\mu}m_\nu\}+\bar{\Sigma}_4\{l_\lambda \bar{m}_{\rho}l_{\mu}\bar{m}_\nu\}\right)\nonumber\\
    &&+\,\left(\Sigma_2\{l_\lambda m_{\rho}k_{\mu}\bar{m}_\nu\}+\bar{\Sigma}_2\{l_\lambda \bar{m}_{\rho}k_{\mu}m_\nu\}\right)\nonumber\\
    &&-\,\bar{\Sigma}_1\left(\{l_\lambda k_{\rho}k_{\mu}m_\nu\}+\{k_\lambda m_{\rho}m_{\mu}\bar{m}_\nu\}\right)-\Sigma_1\left(\{l_\lambda k_{\rho}k_{\mu}\bar{m}_\nu\}+\{k_\lambda \bar{m}_{\rho}\bar{m}_{\mu}m_\nu\}\right)\nonumber\\
    &&+\,\Sigma_3\left(\{l_\lambda k_{\rho}l_{\mu}m_\nu\}-\{l_\lambda m_{\rho}m_{\mu}\bar{m}_\nu\}\right)+\bar{\Sigma}_3\left(\{l_\lambda k_{\rho}l_{\mu}\bar{m}_\nu\}-\{l_\lambda \bar{m}_{\rho}\bar{m}_{\mu}m_\nu\}\right)\,,
    \end{eqnarray}
where
\begin{eqnarray}\label{Sigma0}
  \Sigma_0&=&-\,{}^{(1)}\tilde{W}_{\lambda\rho\mu\nu}l^\lambda m^\rho l^\mu m^\nu\,,\quad\Sigma_1=-\,{}^{(1)}\tilde{W}_{\lambda\rho\mu\nu}l^\lambda k^\rho l^\mu m^\nu\,,\quad\Sigma_2=-\,{}^{(1)}\tilde{W}_{\lambda\rho\mu\nu}l^\lambda m^\rho \bar{m}^\mu k^\nu\,,\\
   \Sigma_3&=&-\,{}^{(1)}\tilde{W}_{\lambda\rho\mu\nu}l^\lambda k^\rho \bar{m}^\mu k^\nu\,,\quad\Sigma_4=-\,{}^{(1)}\tilde{W}_{\lambda\rho\mu\nu}k^\lambda \bar{m}^\rho k^\mu \bar{m}^\nu\,,\label{Sigma4}
\end{eqnarray}
and
\begin{equation}
  \{w_\lambda x_{\rho}y_{\mu}z_\nu\}=w_\lambda x_\rho y_\mu z_\nu-w_\lambda x_\rho z_\mu y_\nu-x_\lambda w_\rho y_\mu z_\nu+x_\lambda w_\rho z_\mu y_\nu+y_\lambda z_\rho w_\mu x_\nu-y_\lambda z_\rho x_\mu w_\nu-z_\lambda y_\rho w_\mu x_\nu+z_\lambda y_\rho x_\mu w_\nu\,.
\end{equation}
From Expression~\eqref{W1id}, it is straightforward to obtain the following identities, which shall provide specific properties for the different algebraic types of the classification:
\begin{align}\label{contractionsids}
    l_{[\sigma}{}^{(1)}\tilde{W}_{\lambda] \rho \mu [\nu}l_{\omega]}l^{\rho}l^{\mu}  &=-\left(\Sigma_0 \bar{m}_{[\sigma}l_{\lambda]}\bar{m}_{[\nu}l_{\omega]}+\bar{\Sigma}_0 m_{[\sigma}l_{\lambda]}m_{[\nu}l_{\omega]}\right)\,,\\ 
    ^{(1)}\tilde{W}_{\lambda \rho \mu [\nu}l_{\omega]}l^{\rho}l^{\mu}&=\left(\Sigma_{0}\bar{m}_{\lambda}-\Sigma_{1}l_{\lambda}\right)\bar{m}_{[\nu}l_{\omega]}+\left(\bar{\Sigma}_{0}m_{\lambda}-\bar{\Sigma}_{1}l_{\lambda}\right)m_{[\nu}l_{\omega]}\,,\\
    ^{(1)}\tilde{W}_{\lambda \rho \mu [\nu}k_{\omega]}k^{\rho}k^{\mu}&=\Sigma_4 m_\lambda m_{[\nu}k_{\omega]}+\bar{\Sigma}_4 \bar{m}_\lambda \bar{m}_{[\nu}k_{\omega]}-\left(\Sigma_3 m_{[\nu}k_{\omega]}+\bar{\Sigma}_3 \bar{m}_{[\nu}k_{\omega]}\right)k_{\lambda}\,,\\
    ^{(1)}\tilde{W}_{\lambda \rho \mu [\nu}l_{\omega]}l^{\mu}&=2\Sigma_0 \bar{m}_{[\lambda}k_{\rho]}\bar{m}_{[\nu}l_{\omega]}+2\bar{\Sigma}_0 m_{[\lambda}k_{\rho]}m_{[\nu}l_{\omega]}+2\Sigma_1(k_{[\lambda}l_{\rho]}-\bar{m}_{[\lambda}m_{\rho]})\bar{m}_{[\nu}l_{\omega]}\nonumber\\
    &+2\bar{\Sigma}_1(k_{[\lambda}l_{\rho]}+\bar{m}_{[\lambda}m_{\rho]})m_{[\nu}l_{\omega]}-2\Sigma_2m_{[\lambda}l_{\rho]}\bar{m}_{[\nu}l_{\omega]}-2\bar{\Sigma}_2\bar{m}_{[\lambda}l_{\rho]}m_{[\nu}l_{\omega]}\,,\\
    ^{(1)}\tilde{W}_{\lambda \rho \mu \nu}l^{\mu}&=2\bigl[\left(\Sigma_{3}m_{[\lambda}l_{\rho]}+\bar{\Sigma}_{3}\bar{m}_{[\lambda}l_{\rho]}\right)l_{\nu}+\left(\Sigma_0\bar{m}_{[\lambda}k_{\rho]}\bar{m}_\nu+\bar{\Sigma}_0m_{[\lambda}k_{\rho]}m_\nu\right)\nonumber\\
    &+\,\Sigma_1\left(k_{[\lambda}l_{\rho]}\bar{m}_\nu-\bar{m}_{[\lambda}m_{\rho]}\bar{m}_\nu-\bar{m}_{[\lambda}k_{\rho]}l_\nu\right)+\bar{\Sigma}_1\left(k_{[\lambda}l_{\rho]}m_\nu+\bar{m}_{[\lambda}m_{\rho]}m_\nu-m_{[\lambda}k_{\rho]}l_\nu\right)\nonumber\\
    &+\left(\Sigma_2-\bar{\Sigma}_2\right)\bar{m}_{[\lambda}m_{\rho]}l_\nu-\left(\Sigma_2+\bar{\Sigma}_2\right)k_{[\lambda}l_{\rho]}l_\nu-\Sigma_2 m_{[\lambda}l_{\rho]}\bar{m}_\nu-\bar{\Sigma}_2\bar{m}_{[\lambda}l_{\rho]}m_\nu\bigr]\,.
\end{align}
In that case, by performing a rotation defined by a complex function $\epsilon$ that keeps the null vector $k^{\mu}$ fixed
\begin{eqnarray}\label{transformation}
    k'_\mu=k_\mu\,,\quad m'_\mu=m_\mu+\epsilon \,k_\mu\,,\quad \bar{m}'_\mu=\bar{m}_\mu+\bar{\epsilon}\,k_\mu\,,\quad l'_\mu=l_\mu+\bar{\epsilon}\,m_\mu+\epsilon\,\bar{m}_\mu+|\epsilon|^{2}k_\mu\,,
\end{eqnarray}
the complex quantities~\eqref{Sigma0}-\eqref{Sigma4} are transformed as
\begin{eqnarray}
\Sigma'_4&=&\Sigma_4\,,\quad \Sigma'_3=\Sigma_3+\epsilon\,\Sigma_4\,,\quad \Sigma'_2=\Sigma_2+2\epsilon\,\Sigma_3+\epsilon^{2}\Sigma_4\,,\\
\Sigma'_1&=&\Sigma_1+3\epsilon\,\Sigma_2+3\epsilon^{2}\Sigma_3+\epsilon^{3}\Sigma_4\,,\\
\Sigma'_0&=&\Sigma_0+4\epsilon\,\Sigma_1+6\epsilon^{2}\Sigma_2+4\epsilon^{3}\Sigma_3+\epsilon^{4}\Sigma_4\,.
\end{eqnarray}
Thereby, if we demand that $\Sigma'_0=0$ under the transformation~\eqref{transformation}, the corresponding multiplicities of the roots of the quartic polynomial
\begin{equation}\label{quarticpolW1}
    \Sigma_0+4\epsilon\,\Sigma_1+6\epsilon^{2}\Sigma_2+4\epsilon^{3}\Sigma_3+\epsilon^{4}\Sigma_4=0\,,
\end{equation}
lead to an algebraic classification which is equivalent to the one obtained from the eigenvalue problem~\eqref{eigenW1}\footnote{For simplicity, we omit the tilde in the notation for each equivalence.}:
\begin{eqnarray}
    l_{[\sigma}{}^{(1)}\tilde{W}_{\lambda] \rho \mu [\nu}l_{\omega]}l^{\rho}l^{\mu}=0 &\iff& \Sigma_0=0\,,\\
    ^{(1)}\tilde{W}_{\lambda \rho \mu [\nu}l_{\omega]}l^{\rho}l^{\mu}=0 &\iff& \Sigma_0=\Sigma_1=0\,,\\
    ^{(1)}\tilde{W}_{\lambda \rho \mu [\nu}k_{\omega]}k^{\rho}k^{\mu}={}^{(1)}\tilde{W}_{\lambda \rho \mu [\nu}l_{\omega]}l^{\rho}l^{\mu}=0 &\iff& \Sigma_0=\Sigma_1=\Sigma_3=\Sigma_4=0\,,\\
    ^{(1)}\tilde{W}_{\lambda \rho \mu [\nu}l_{\omega]}l^{\mu}=0 &\iff& \Sigma_0=\Sigma_1=\Sigma_2=0\,,\\
    ^{(1)}\tilde{W}_{\lambda \rho \mu \nu}l^{\mu}=0 &\iff& \Sigma_0=\Sigma_1=\Sigma_2=\Sigma_3=0\,.
\end{eqnarray}

The results arising from these two equivalent procedures can then be summarised in Table~\ref{tab:Algebraictypes1}.

\begin{table*}
\begin{center}
    \begin{tabular}{| c | c | c | c|}
    \hline
    Algebraic type & Segre characteristic & Invariants & Constraints with the principal null directions \\ \hline
    
    $I$ & $[1\,1\,1]$  & $I^3 \neq 12 J^2$ & $\text{No further constraints}$ \\ \hline
    $II$ & $[2\,1]$  & $I^3 = 12 J^2$ & $^{(1)}\tilde{W}_{\lambda \rho \mu [\nu}l_{\omega]}l^{\rho}l^{\mu}=0$ \\ \hline
    $D$ & $[(1\,1)\,1]$  & $I^3 = 12 J^2$ & $^{(1)}\tilde{W}_{\lambda \rho \mu [\nu}k_{\omega]}k^{\rho}k^{\mu}={}^{(1)}\tilde{W}_{\lambda \rho \mu [\nu}l_{\omega]}l^{\rho}l^{\mu}=0$ \\ \hline
    $III$ & $[3]$  & $I=J=0$ & $^{(1)}\tilde{W}_{\lambda \rho \mu [\nu}l_{\omega]}l^{\mu}=0$ \\ \hline
    $N$ & $[(2\,1)]$  & $I=J=0$ & $^{(1)}\tilde{W}_{\lambda \rho \mu \nu}l^{\mu}=0$ \\ \hline
    $O$ & $[-]$  & $I=J=0$ & $^{(1)}\tilde{W}_{\lambda \rho \mu \nu}=0$ \\ \hline
    \end{tabular}
\end{center}
\caption{Algebraic types for the tensor $^{(1)}\tilde{W}_{\lambda\rho\mu\nu}$.}
\label{tab:Algebraictypes1}
\end{table*}

\subsection{Algebraic classification of ${\nearrow\!\!\!\!\!\!\!\tilde{R}}^{(T)}_{\lambda[\rho\mu\nu]}$ and ${\nearrow\!\!\!\!\!\!\!\tilde{R}}_{(\mu\nu)}$}

For the classification of the antisymmetrised traceless and pseudotraceless piece of the curvature tensor, it is worthwhile to note the number of independent components for this piece is $9$, in virtue of the relations
\begin{align}
    {\nearrow\!\!\!\!\!\!\!\tilde{R}}^{(T)}_{\hat{3}[\hat{1}\hat{2}\hat{3}]}&=-\,{\nearrow\!\!\!\!\!\!\!\tilde{R}}^{(T)}_{\hat{0}[\hat{0}\hat{2}\hat{1}]}\,, \quad {\nearrow\!\!\!\!\!\!\!\tilde{R}}^{(T)}_{\hat{2}[\hat{1}\hat{2}\hat{3}]}=-\,{\nearrow\!\!\!\!\!\!\!\tilde{R}}^{(T)}_{\hat{0}[\hat{0}\hat{1}\hat{3}]}\,, \quad {\nearrow\!\!\!\!\!\!\!\tilde{R}}^{(T)}_{\hat{1}[\hat{1}\hat{2}\hat{3}]}=-\,{\nearrow\!\!\!\!\!\!\!\tilde{R}}^{(T)}_{\hat{0}[\hat{0}\hat{3}\hat{2}]}\,,\\
    {\nearrow\!\!\!\!\!\!\!\tilde{R}}^{(T)}_{\hat{3}[\hat{0}\hat{1}\hat{3}]}&={\nearrow\!\!\!\!\!\!\!\tilde{R}}^{(T)}_{\hat{2}[\hat{0}\hat{2}\hat{1}]}\,, \quad {\nearrow\!\!\!\!\!\!\!\tilde{R}}^{(T)}_{\hat{3}[\hat{0}\hat{2}\hat{3}]}=-\,{\nearrow\!\!\!\!\!\!\!\tilde{R}}^{(T)}_{\hat{1}[\hat{0}\hat{2}\hat{1}]}\,, \quad {\nearrow\!\!\!\!\!\!\!\tilde{R}}^{(T)}_{\hat{2}[\hat{0}\hat{2}\hat{3}]}=-\,{\nearrow\!\!\!\!\!\!\!\tilde{R}}^{(T)}_{\hat{1}[\hat{0}\hat{1}\hat{3}]}\,,\\
        {\nearrow\!\!\!\!\!\!\!\tilde{R}}^{(T)}_{\hat{3}[\hat{0}\hat{1}\hat{2}]}&={\nearrow\!\!\!\!\!\!\!\tilde{R}}^{(T)}_{\hat{1}[\hat{0}\hat{3}\hat{2}]}+{\nearrow\!\!\!\!\!\!\!\tilde{R}}^{(T)}_{\hat{2}[\hat{0}\hat{1}\hat{3}]}-{\nearrow\!\!\!\!\!\!\!\tilde{R}}^{(T)}_{\hat{0}[\hat{1}\hat{3}\hat{2}]}\,.
\end{align}
Without any loss of generality, we consider $\{{\nearrow\!\!\!\!\!\!\!\tilde{R}}^{(T)}_{\hat{0}[\hat{1}\hat{3}\hat{2}]},{\nearrow\!\!\!\!\!\!\!\tilde{R}}^{(T)}_{\hat{0}[\hat{0}\hat{2}\hat{1}]},{\nearrow\!\!\!\!\!\!\!\tilde{R}}^{(T)}_{\hat{0}[\hat{0}\hat{1}\hat{3}]},{\nearrow\!\!\!\!\!\!\!\tilde{R}}^{(T)}_{\hat{0}[\hat{0}\hat{3}\hat{2}]},{\nearrow\!\!\!\!\!\!\!\tilde{R}}^{(T)}_{\hat{1}[\hat{0}\hat{2}\hat{1}]},{\nearrow\!\!\!\!\!\!\!\tilde{R}}^{(T)}_{\hat{1}[\hat{0}\hat{1}\hat{3}]},{\nearrow\!\!\!\!\!\!\!\tilde{R}}^{(T)}_{\hat{1}[\hat{0}\hat{3}\hat{2}]},{\nearrow\!\!\!\!\!\!\!\tilde{R}}^{(T)}_{\hat{2}[\hat{0}\hat{2}\hat{1}]},{\nearrow\!\!\!\!\!\!\!\tilde{R}}^{(T)}_{\hat{2}[\hat{0}\hat{1}\hat{3}]}\}$ as the set of independent components for the aforementioned piece, which can be encoded in a second order symmetric and traceless tensor
\begin{equation}\label{Matrix_Antisymmetrised_Curvature}
    {\nearrow\!\!\!\!\!\!\!M}_{a b}=\frac{1}{6}\varepsilon_{(a}{}^{c d e}{\nearrow\!\!\!\!\!\!\!\tilde{R}}^{(T)}_{b)[c d e]}\,,
\end{equation}
as follows:
\begin{eqnarray}
{\nearrow\!\!\!\!\!\!\!M}_{a b} &=& \left(
\begin{array}{cccc}
{\nearrow\!\!\!\!\!\!\!\tilde{R}}^{(T)}_{\hat{0}[\hat{1}\hat{3}\hat{2}]} & {\nearrow\!\!\!\!\!\!\!\tilde{R}}^{(T)}_{\hat{0}[\hat{0}\hat{3}\hat{2}]} & {\nearrow\!\!\!\!\!\!\!\tilde{R}}^{(T)}_{\hat{0}[\hat{0}\hat{1}\hat{3}]} & {\nearrow\!\!\!\!\!\!\!\tilde{R}}^{(T)}_{\hat{0}[\hat{0}\hat{2}\hat{1}]} \\
{\nearrow\!\!\!\!\!\!\!\tilde{R}}^{(T)}_{\hat{0}[\hat{0}\hat{3}\hat{2}]} & {\nearrow\!\!\!\!\!\!\!\tilde{R}}^{(T)}_{\hat{1}[\hat{0}\hat{3}\hat{2}]} & {\nearrow\!\!\!\!\!\!\!\tilde{R}}^{(T)}_{\hat{1}[\hat{0}\hat{1}\hat{3}]} & {\nearrow\!\!\!\!\!\!\!\tilde{R}}^{(T)}_{\hat{1}[\hat{0}\hat{2}\hat{1}]} \\
{\nearrow\!\!\!\!\!\!\!\tilde{R}}^{(T)}_{\hat{0}[\hat{0}\hat{1}\hat{3}]} & {\nearrow\!\!\!\!\!\!\!\tilde{R}}^{(T)}_{\hat{1}[\hat{0}\hat{1}\hat{3}]} & {\nearrow\!\!\!\!\!\!\!\tilde{R}}^{(T)}_{\hat{2}[\hat{0}\hat{1}\hat{3}]} & {\nearrow\!\!\!\!\!\!\!\tilde{R}}^{(T)}_{\hat{2}[\hat{0}\hat{2}\hat{1}]} \\
{\nearrow\!\!\!\!\!\!\!\tilde{R}}^{(T)}_{\hat{0}[\hat{0}\hat{2}\hat{1}]} & {\nearrow\!\!\!\!\!\!\!\tilde{R}}^{(T)}_{\hat{1}[\hat{0}\hat{2}\hat{1}]} & {\nearrow\!\!\!\!\!\!\!\tilde{R}}^{(T)}_{\hat{2}[\hat{0}\hat{2}\hat{1}]} & {\nearrow\!\!\!\!\!\!\!\tilde{R}}^{(T)}_{\hat{0}[\hat{1}\hat{3}\hat{2}]}-{\nearrow\!\!\!\!\!\!\!\tilde{R}}^{(T)}_{\hat{1}[\hat{0}\hat{3}\hat{2}]}-{\nearrow\!\!\!\!\!\!\!\tilde{R}}^{(T)}_{\hat{2}[\hat{0}\hat{1}\hat{3}]}
\end{array} \right)\,.
\end{eqnarray}

Likewise, we include the $9$ independent components $\{{\nearrow\!\!\!\!\!\!\!\tilde{R}}_{(\hat{0}\hat{0})},{\nearrow\!\!\!\!\!\!\!\tilde{R}}_{(\hat{0}\hat{1})},{\nearrow\!\!\!\!\!\!\!\tilde{R}}_{(\hat{0}\hat{2})},{\nearrow\!\!\!\!\!\!\!\tilde{R}}_{(\hat{0}\hat{3})},{\nearrow\!\!\!\!\!\!\!\tilde{R}}_{(\hat{1}\hat{1})},{\nearrow\!\!\!\!\!\!\!\tilde{R}}_{(\hat{1}\hat{2})},{\nearrow\!\!\!\!\!\!\!\tilde{R}}_{(\hat{1}\hat{3})},{\nearrow\!\!\!\!\!\!\!\tilde{R}}_{(\hat{2}\hat{2})},{\nearrow\!\!\!\!\!\!\!\tilde{R}}_{(\hat{2}\hat{3})}\}$ in the symmetric and traceless tensor
\begin{eqnarray}
{\nearrow\!\!\!\!\!\!\!N}_{a b} &=& \left(
\begin{array}{cccc}
{\nearrow\!\!\!\!\!\!\!\tilde{R}}_{(\hat{0}\hat{0})} & {\nearrow\!\!\!\!\!\!\!\tilde{R}}_{(\hat{0}\hat{1})} & {\nearrow\!\!\!\!\!\!\!\tilde{R}}_{(\hat{0}\hat{2})} & {\nearrow\!\!\!\!\!\!\!\tilde{R}}_{(\hat{0}\hat{3})} \\
{\nearrow\!\!\!\!\!\!\!\tilde{R}}_{(\hat{1}\hat{0})} & {\nearrow\!\!\!\!\!\!\!\tilde{R}}_{(\hat{1}\hat{1})} & {\nearrow\!\!\!\!\!\!\!\tilde{R}}_{(\hat{1}\hat{2})} & {\nearrow\!\!\!\!\!\!\!\tilde{R}}_{(\hat{1}\hat{3})} \\
{\nearrow\!\!\!\!\!\!\!\tilde{R}}_{(\hat{2}\hat{0})} & {\nearrow\!\!\!\!\!\!\!\tilde{R}}_{(\hat{2}\hat{1})} & {\nearrow\!\!\!\!\!\!\!\tilde{R}}_{(\hat{2}\hat{2})} & {\nearrow\!\!\!\!\!\!\!\tilde{R}}_{(\hat{2}\hat{3})} \\
{\nearrow\!\!\!\!\!\!\!\tilde{R}}_{(\hat{3}\hat{0})} & {\nearrow\!\!\!\!\!\!\!\tilde{R}}_{(\hat{3}\hat{1})} & {\nearrow\!\!\!\!\!\!\!\tilde{R}}_{(\hat{3}\hat{2})} & {\nearrow\!\!\!\!\!\!\!\tilde{R}}_{(\hat{0}\hat{0})}-{\nearrow\!\!\!\!\!\!\!\tilde{R}}_{(\hat{1}\hat{1})}-{\nearrow\!\!\!\!\!\!\!\tilde{R}}_{(\hat{2}\hat{2})}
\end{array} \right)\,.
\end{eqnarray}

Accordingly, both tensors ${\nearrow\!\!\!\!\!\!\!\tilde{R}}^{(T)}_{\lambda[\rho\mu\nu]}$ and ${\nearrow\!\!\!\!\!\!\!\tilde{R}}_{(\mu\nu)}$ obey the same class of algebraic classification, which can be directly derived from the eigenvalue equations
\begin{align}
    {\nearrow\!\!\!\!\!\!\!M}^{a}{}_{b}v^{b}&=\lambda v^{a}\,,\\
    {\nearrow\!\!\!\!\!\!\!N}^{a}{}_{b}w^{b}&=\sigma w^{a}\,.
\end{align}

The respective characteristic polynomials can then be determined from the invariants
\begin{align}
    U_1&={\nearrow\!\!\!\!\!\!\!M}^{a}{}_{b}{\nearrow\!\!\!\!\!\!\!M}^{b}{}_{a}\,, \quad V_1={\nearrow\!\!\!\!\!\!\!M}^{a}{}_{b}{\nearrow\!\!\!\!\!\!\!M}^{b}{}_{c}{\nearrow\!\!\!\!\!\!\!M}^{c}{}_{a}\,, \quad W_1={\nearrow\!\!\!\!\!\!\!M}^{a}{}_{b}{\nearrow\!\!\!\!\!\!\!M}^{b}{}_{c}{\nearrow\!\!\!\!\!\!\!M}^{c}{}_{d}{\nearrow\!\!\!\!\!\!\!M}^{d}{}_{a}\,,\\
    U_2&={\nearrow\!\!\!\!\!\!\!N}^{a}{}_{b}{\nearrow\!\!\!\!\!\!\!N}^{b}{}_{a}\,, \quad V_2={\nearrow\!\!\!\!\!\!\!N}^{a}{}_{b}{\nearrow\!\!\!\!\!\!\!N}^{b}{}_{c}{\nearrow\!\!\!\!\!\!\!N}^{c}{}_{a}\,, \quad W_2={\nearrow\!\!\!\!\!\!\!N}^{a}{}_{b}{\nearrow\!\!\!\!\!\!\!N}^{b}{}_{c}{\nearrow\!\!\!\!\!\!\!N}^{c}{}_{d}{\nearrow\!\!\!\!\!\!\!N}^{d}{}_{a}\,,
\end{align}
yielding
\begin{align}
    \lambda^4-\frac{U_1}{2}\lambda^2-\frac{V_1}{3}\lambda+\frac{1}{8}\left(U_1^{2}-2W_1\right)&=0\,,\\
    \sigma^4-\frac{U_2}{2}\sigma^2-\frac{V_2}{3}\sigma+\frac{1}{8}\left(U_2^{2}-2W_2\right)&=0\,.
\end{align}

In particular, it is well-known that the corresponding multiplicities of the roots of these characteristic equations are determined by certain combinations of signs given by the subsequent invariants~\cite{bona1992intrinsic,Senovilla:1999fa}:
\begin{align}
    U_{1}^{*}&=W_{1}^{*\,3}-\left[3U_{1}W_{1}^{*}+4\left(3V_{1}^{2}-U_{1}^{3}\right)\right]^2\,, \quad V_{1}^{*}=2U_{1}-|W_{1}^{*}|^{1/2} \,, \quad W_{1}^{*}=7U_{1}^{2}-12W_{1}\,,\\
    U_{2}^{*}&=W_{2}^{*\,3}-\left[3U_{1}W_{2}^{*}+4\left(3V_{2}^{2}-U_{2}^{3}\right)\right]^2\,, \quad V_{2}^{*}=2U_{2}-|W_{2}^{*}|^{1/2} \,, \quad W_{2}^{*}=7U_{2}^{2}-12W_{2}\,,
\end{align}
leading to the $15$ different algebraic types described in Tables~\ref{tab:Algebraictypes2} and~\ref{tab:Algebraictypes3}.

\begin{table*}
\begin{center}
    \begin{tabular}{| c | c |}
    \hline
    Segre characteristic & Invariants \\ \hline
    $[1,1\,1\,1]$  & $U_{1}^{*}, V_{1}^{*} > 0$  \\ \hline
    $[Z\,\bar{Z}\,1\,1]$  & $U_{1}^{*} < 0$  \\ \hline
    $[Z\,\bar{Z}\,(1\,1)]$  & $U_{1}^{*} = 0\,, \,\, V_{1}^{*} < 0\,, \,\,  W_{1}^{*} > 0$  \\ \hline
    $[2\,1\,1]$  & $U_{1}^{*} = 0\,, \,\, V_{1}^{*} > 0\,, \,\,  W_{1}^{*} > 0$  \\ \hline
    $[1,1\,(1\,1)]$  & $U_{1}^{*} = 0\,, \,\, V_{1}^{*} > 0\,, \,\,  W_{1}^{*} > 0$  \\ \hline
    $[(1,1)1\,1]$  & $U_{1}^{*} = 0\,, \,\, V_{1}^{*} > 0\,, \,\,  W_{1}^{*} > 0$  \\ \hline
    $[3\,1]$  & $U_{1}^{*}=W_{1}^{*}=0\,, \,\, V_{1}^{*} > 0$  \\ \hline
    $[(2\,1)\,1]$  & $U_{1}^{*}=W_{1}^{*}=0\,, \,\, V_{1}^{*} > 0$  \\ \hline
    $[(1,1\,1)\,1]$  & $U_{1}^{*}=W_{1}^{*}=0\,, \,\, V_{1}^{*} > 0$  \\ \hline
    $[1,(1\,1\,1)]$  & $U_{1}^{*}=W_{1}^{*}=0\,, \,\, V_{1}^{*} > 0$  \\ \hline
    $[2\,(1\,1)]$  & $U_{1}^{*}=V_{1}^{*}=0\,, \,\,  W_{1}^{*} > 0$  \\ \hline
    $[(1,1)\,(1\,1)]$  & $U_{1}^{*}=V_{1}^{*}=0\,, \,\,  W_{1}^{*} > 0$  \\ \hline
    $[(3\,1)]$  & $U_{1}^{*}=V_{1}^{*}=W_{1}^{*}=0$  \\ \hline
    $[(2\,1\,1)]$  & $U_{1}^{*}=V_{1}^{*}=W_{1}^{*}=0$  \\ \hline
    $[(1,1\,1\,1)]$  & $U_{1}^{*}=V_{1}^{*}=W_{1}^{*}=0$  \\ \hline
    \end{tabular}
\end{center}
\caption{Algebraic types for the tensor ${\nearrow\!\!\!\!\!\!\!\tilde{R}}^{(T)}_{\lambda[\rho\mu\nu]}$.}
\label{tab:Algebraictypes2}
\end{table*}

\begin{table*}
\begin{center}
    \begin{tabular}{| c | c |}
    \hline
    Segre characteristic & Invariants \\ \hline
    $[1,1\,1\,1]$  & $U_{2}^{*}, V_{2}^{*} > 0$  \\ \hline
    $[Z\,\bar{Z}\,1\,1]$  & $U_{2}^{*} < 0$  \\ \hline
    $[Z\,\bar{Z}\,(1\,1)]$  & $U_{2}^{*} = 0\,, \,\, V_{2}^{*} < 0\,, \,\,  W_{2}^{*} > 0$  \\ \hline
    $[2\,1\,1]$  & $U_{2}^{*} = 0\,, \,\, V_{2}^{*} > 0\,, \,\,  W_{2}^{*} > 0$  \\ \hline
    $[1,1\,(1\,1)]$  & $U_{2}^{*} = 0\,, \,\, V_{2}^{*} > 0\,, \,\,  W_{2}^{*} > 0$  \\ \hline
    $[(1,1)1\,1]$  & $U_{2}^{*} = 0\,, \,\, V_{2}^{*} > 0\,, \,\,  W_{2}^{*} > 0$  \\ \hline
    $[3\,1]$  & $U_{2}^{*}=W_{2}^{*}=0\,, \,\, V_{2}^{*} > 0$  \\ \hline
    $[(2\,1)\,1]$  & $U_{2}^{*}=W_{2}^{*}=0\,, \,\, V_{2}^{*} > 0$  \\ \hline
    $[(1,1\,1)\,1]$  & $U_{2}^{*}=W_{2}^{*}=0\,, \,\, V_{2}^{*} > 0$  \\ \hline
    $[1,(1\,1\,1)]$  & $U_{2}^{*}=W_{2}^{*}=0\,, \,\, V_{2}^{*} > 0$  \\ \hline
    $[2\,(1\,1)]$  & $U_{2}^{*}=V_{2}^{*}=0\,, \,\,  W_{2}^{*} > 0$  \\ \hline
    $[(1,1)\,(1\,1)]$  & $U_{2}^{*}=V_{2}^{*}=0\,, \,\,  W_{2}^{*} > 0$  \\ \hline
    $[(3\,1)]$  & $U_{2}^{*}=V_{2}^{*}=W_{2}^{*}=0$  \\ \hline
    $[(2\,1\,1)]$  & $U_{2}^{*}=V_{2}^{*}=W_{2}^{*}=0$  \\ \hline
    $[(1,1\,1\,1)]$  & $U_{2}^{*}=V_{2}^{*}=W_{2}^{*}=0$  \\ \hline
    \end{tabular}
\end{center}
\caption{Algebraic types for the tensor ${\nearrow\!\!\!\!\!\!\!\tilde{R}}_{(\mu\nu)}$.}
\label{tab:Algebraictypes3}
\end{table*}

\subsection{Algebraic classification of $\tilde{R}^{(T)}_{[\mu\nu]}$ and $\tilde{R}^{\lambda}{}_{\lambda\mu\nu}$}\label{Algebraic_Class3}

For the algebraic classification of $\tilde{R}^{(T)}_{[\mu\nu]}$ and $\tilde{R}^{\lambda}{}_{\lambda\mu\nu}$, we collect their components in the symmetric spinors
\begin{align}
    \Xi_{AB}&=\frac{1}{2}\varepsilon^{\dot{C}\dot{D}}\tau^{\mu}{}_{A\dot{C}}\tau^{\nu}{}_{B\dot{D}}\tilde{R}^{(T)}_{[\mu\nu]}\,,\\
    \Upsilon_{AB}&=\frac{1}{2}\varepsilon^{\dot{C}\dot{D}}\tau^{\mu}{}_{A\dot{C}}\tau^{\nu}{}_{B\dot{D}}\tilde{R}^{\lambda}{}_{\lambda\mu\nu}\,.
\end{align}
Then, both quantities can be classified according to the eigenvalue equations for two different eigenspinors $\xi^{A}$ and $\zeta^{A}$:
\begin{align}
    \Xi^{A}{}_{B}\xi^{B}&=\lambda\xi^{A}\,,\\
    \Upsilon^{A}{}_{B}\zeta^{B}&=\sigma\zeta^{A}\,,
\end{align}
whose eigenvalues are given by the invariants
\begin{equation}
    X = \Xi_{AB}\Xi^{AB} \,, \quad Y = \Upsilon_{AB}\Upsilon^{AB}\,,
\end{equation}
as follows:
\begin{equation}
    \lambda=\pm\,\sqrt{-\,\frac{X}{2}}\,,\quad\sigma=\pm\,\sqrt{-\,\frac{Y}{2}}\,.
\end{equation}

In terms of the null vectors, the tensors $\tilde{R}^{(T)}_{[\mu\nu]}$ and $\tilde{R}^{\lambda}{}_{\lambda\mu\nu}$ satisfy the identities
\begin{align}
    \tilde{R}^{(T)}_{[\mu\nu]}&=2\left[\Omega_{2}k_{[\mu}m_{\nu]}+\bar{\Omega}_{2}k_{[\mu}\bar{m}_{\nu]}-\Omega_{0}l_{[\mu}\bar{m}_{\nu]}-\bar{\Omega}_{0}l_{[\mu}m_{\nu]}-\left(\Omega_{1}+\bar{\Omega}_{1}\right)k_{[\mu}l_{\nu]}+\left(\Omega_{1}-\bar{\Omega}_{1}\right)m_{[\mu}\bar{m}_{\nu]}\right]\,,\\
    \tilde{R}^{\lambda}{}_{\lambda\mu\nu}&=2\left[\Pi_{2}k_{[\mu}m_{\nu]}+\bar{\Pi}_{2}k_{[\mu}\bar{m}_{\nu]}-\Pi_{0}l_{[\mu}\bar{m}_{\nu]}-\bar{\Pi}_{0}l_{[\mu}m_{\nu]}-\left(\Pi_{1}+\bar{\Pi}_{1}\right)k_{[\mu}l_{\nu]}+\left(\Pi_{1}-\bar{\Pi}_{1}\right)m_{[\mu}\bar{m}_{\nu]}\right]\,,
\end{align}
where
\begin{align}
    \Omega_{0}&=k^{[\mu}m^{\nu]}\tilde{R}^{(T)}_{[\mu\nu]}\,, \quad \Omega_{1}=\frac{1}{2}\Bigl(k^{[\mu}l^{\nu]}-m^{[\mu}\bar{m}^{\nu]}\Bigr)\tilde{R}^{(T)}_{[\mu\nu]}\,, \quad \Omega_{2}=-\,l^{[\mu}\bar{m}^{\nu]}\tilde{R}^{(T)}_{[\mu\nu]}\,,\label{Omega}\\
    \Pi_{0}&=k^{[\mu}m^{\nu]}\tilde{R}^{\lambda}{}_{\lambda\mu\nu}\,, \quad \Pi_{1}=\frac{1}{2}\Bigl(k^{[\mu}l^{\nu]}-m^{[\mu}\bar{m}^{\nu]}\Bigr)\tilde{R}^{\lambda}{}_{\lambda\mu\nu}\,, \quad \Pi_{2}=-\,l^{[\mu}\bar{m}^{\nu]}\tilde{R}^{\lambda}{}_{\lambda\mu\nu}\,,\label{Pi}
\end{align}
are six complex quantities encoding the independent components of these tensors. Thereby, a rotation of the form~\eqref{transformation} transforms the aforementioned quantities as
\begin{eqnarray}
    \Omega_0'&=&\Omega_0\,,\quad \Omega_1'=\Omega_1+\bar{\epsilon}\,\Omega_0\,,\quad \Omega_2'=\Omega_2+2\bar{\epsilon}\,\Omega_1+\bar{\epsilon}^2\Omega_0\,,\\
    \Pi_0'&=&\Pi_0\,,\quad \Pi_1'=\Pi_1+\bar{\epsilon}\,\Pi_0\,,\quad \Pi_2'=\Pi_2+2\bar{\epsilon}\,\Pi_1+\bar{\epsilon}^2\Pi_0\,,
\end{eqnarray}
which leads to an equivalent classification for the different multiplicities of the roots of the quadratic polynomials
\begin{align}
    \Omega_2+2\bar{\epsilon}\,\Omega_1+\bar{\epsilon}^2\Omega_0&=0\,,\label{quadraticpolOmega}\\
    \Pi_2+2\bar{\epsilon}\,\Pi_1+\bar{\epsilon}^2\Pi_0&=0\,.\label{quadraticpolPi}
\end{align}
Specifically, from the contractions
\begin{align}
    \bigl(\tilde{R}^{(T)}_{[\mu\nu]}l_{\lambda}-\tilde{R}^{(T)}_{[\mu\lambda]}l_{\nu}\bigr)l^{\mu}&=2\left[\Omega_2\left(k_{[\mu}m_{\nu]}l^\mu l_{\lambda} -k_{[\mu}m_{\lambda]}l^\mu l_{\nu}\right)+\bar{\Omega}_2\left(k_{[\mu}\bar{m}_{\nu]}l^\mu l_{\lambda} -k_{[\mu}\bar{m}_{\lambda]}l^\mu l_{\nu}\right)\right]\,,\\
    \bigl(\tilde{R}^{\sigma}{}_{\sigma\mu\nu}l_{\lambda}-\tilde{R}^{\sigma}{}_{\sigma\mu\lambda}l_{\nu}\bigr)l^{\mu}&=2\left[\Pi_2\left(k_{[\mu}m_{\nu]}l^\mu l_{\lambda} -k_{[\mu}m_{\lambda]}l^\mu l_{\nu}\right)+\bar{\Pi}_2\left(k_{[\mu}\bar{m}_{\nu]}l^\mu l_{\lambda} -k_{[\mu}\bar{m}_{\lambda]}l^\mu l_{\nu}\right)\right]\,,\\
    \bigl(\tilde{R}^{(T)}_{[\mu\nu]}k_{\lambda}-\tilde{R}^{(T)}_{[\mu\lambda]}k_{\nu}\bigr)k^{\mu}&=-\,2\left[\Omega_0\left(l_{[\mu}\bar{m}_{\nu]}k^\mu k_{\lambda} -l_{[\mu}\bar{m}_{\lambda]}k^\mu k_{\nu}\right)+\bar{\Omega}_0\left(l_{[\mu}m_{\nu]}k^\mu k_{\lambda} -l_{[\mu}m_{\lambda]}k^\mu k_{\nu}\right)\right]\,,\\
    \bigl(\tilde{R}^{\sigma}{}_{\sigma\mu\nu}k_{\lambda}-\tilde{R}^{\sigma}{}_{\sigma\mu\lambda}k_{\nu}\bigr)k^{\mu}&=-\,2\left[\Pi_0\left(l_{[\mu}\bar{m}_{\nu]}k^\mu k_{\lambda} -l_{[\mu}\bar{m}_{\lambda]}k^\mu k_{\nu}\right)+\bar{\Pi}_0\left(l_{[\mu}m_{\nu]}k^\mu k_{\lambda} -l_{[\mu}m_{\lambda]}k^\mu k_{\nu}\right)\right]\,,\\
    \tilde{R}^{(T)}_{[\mu\nu]}l^\mu&= 2\left[\Omega_2 k_{[\mu}m_{\nu]}+\bar{\Omega}_2 k_{[\mu}\bar{m}_{\nu]}-\left(\Omega_1+\bar{\Omega}_1\right)k_{[\mu}l_{\nu]}\right]l^\mu\,,\\
    \tilde{R}^{\sigma}{}_{\sigma\mu\nu}l^\mu&= 2\left[\Pi_2 k_{[\mu}m_{\nu]}+\bar{\Pi}_2 k_{[\mu}\bar{m}_{\nu]}-\left(\Pi_1+\bar{\Pi}_1\right)k_{[\mu}l_{\nu]}\right]l^\mu\,,
\end{align}
it is clear to obtain the following set of equivalences:
\begin{eqnarray}
    \bigl(\tilde{R}^{(T)}_{[\mu\nu]}l_{\lambda}-\tilde{R}^{(T)}_{[\mu\lambda]}l_{\nu}\bigr)l^{\mu}=0 &\iff& \Omega_2 = 0\,,\\
    \bigl(\tilde{R}^{\sigma}{}_{\sigma\mu\nu}l_{\lambda}-\tilde{R}^{\sigma}{}_{\sigma\mu\lambda}l_{\nu}\bigr)l^{\mu}=0 &\iff& \Pi_2 = 0\,,\\
    \bigl(\tilde{R}^{(T)}_{[\mu\nu]}k_{\lambda}-\tilde{R}^{(T)}_{[\mu\lambda]}k_{\nu}\bigr)k^{\mu}=0 &\iff& \Omega_0 = 0\,,\\
    \bigl(\tilde{R}^{\sigma}{}_{\sigma\mu\nu}k_{\lambda}-\tilde{R}^{\sigma}{}_{\sigma\mu\lambda}k_{\nu}\bigr)k^{\mu}=0 &\iff& \Pi_0 = 0\,,\\
    \tilde{R}^{(T)}_{[\mu\nu]}l^\mu=0 &\iff& \Omega_1 = \Omega_2 = 0\,,\\
    \tilde{R}^{\sigma}{}_{\sigma\mu\nu}l^\mu=0 &\iff& \Pi_1 = \Pi_2 = 0\,,
\end{eqnarray}
which provide the conditions
\begin{align}
    \bigl(\tilde{R}^{(T)}_{[\mu\nu]}k_{\lambda}-\tilde{R}^{(T)}_{[\mu\lambda]}k_{\nu}\bigr)k^{\mu}&=\bigl(\tilde{R}^{(T)}_{[\mu\nu]}l_{\lambda}-\tilde{R}^{(T)}_{[\mu\lambda]}l_{\nu}\bigr)l^{\mu}=0 \,,\label{typeDricci}\\
    \bigl(\tilde{R}^{\sigma}{}_{\sigma\mu\nu}k_{\lambda}-\tilde{R}^{\sigma}{}_{\sigma\mu\lambda}k_{\nu}\bigr)k^{\mu}&=\bigl(\tilde{R}^{\sigma}{}_{\sigma\mu\nu}l_{\lambda}-\tilde{R}^{\sigma}{}_{\sigma\mu\lambda}l_{\nu}\bigr)l^{\mu}=0\,,
\end{align}
for the null vectors $k^{\mu}$ and $l^{\mu}$ to be principal null directions of the tensors $\tilde{R}^{(T)}_{[\mu\nu]}$ and $\tilde{R}^{\lambda}{}_{\lambda\mu\nu}$, as well as the different algebraic types of the classification, which is shown in Tables~\ref{tab:Algebraictypes4} and~\ref{tab:Algebraictypes5}.

\begin{table*}
\begin{center}
    \begin{tabular}{| c | c | c | c|}
    \hline
    Algebraic type & Segre characteristic & Invariants & Constraints with the principal null directions \\ \hline
    $I$ & $[1\,1]$  & $X \neq 0$ & $\text{No further constraints}$ \\ \hline
    $N$ & $[2]$ & $X = 0$ & $\tilde{R}^{(T)}_{[\mu\nu]}l^{\mu}=0$ \\ \hline
    $O$ & $[-]$ & $X = 0$  & $\tilde{R}^{(T)}_{[\mu\nu]}=0$ \\ \hline
    \end{tabular}
\end{center}
\caption{Algebraic types for the tensor $\tilde{R}^{(T)}_{[\mu\nu]}$.}
\label{tab:Algebraictypes4}
\end{table*}
\begin{table*}
\begin{center}
    \begin{tabular}{| c | c | c | c|}
    \hline
    Algebraic type & Segre characteristic & Invariants & Constraints with the principal null directions \\ \hline
    $I$ & $[1\,1]$  & $Y \neq 0$ & $\text{No further constraints}$ \\ \hline
    $N$ & $[2]$ & $Y = 0$ & $\tilde{R}^{\lambda}{}_{\lambda\mu\nu}l^{\mu}=0$ \\ \hline
    $O$ & $[-]$ & $Y = 0$ & $\tilde{R}^{\lambda}{}_{\lambda\mu\nu}=0$ \\ \hline
    \end{tabular}
\end{center}
\caption{Algebraic types for the tensor $\tilde{R}^{\lambda}{}_{\lambda\mu\nu}$.}
\label{tab:Algebraictypes5}
\end{table*}

\section{Application to stationary and axisymmetric space-times}\label{sec:application}

The geometry of stationary and axisymmetric space-times underlays from two Killing vectors $\partial_{t}$ and $\partial_{\varphi}$, which generate time translations and rotations around a symmetry axis. In particular, the latter defines a regular two-dimensional timelike surface of fixed points where it vanishes and provides a metric structure which is invariant under the action of the rotation group $SO(2)$~\cite{Stephani:2003tm,Griffiths:2009dfa}. Given the fact that these two Killing vectors are not mutually orthogonal, the most general stationary and axisymmetric space-time satisfying circularity conditions can then be expressed with only one nonvanishing off-diagonal component $g_{t\varphi}$~\cite{hartle1967variational}:
\begin{equation}\label{axi_line}
    ds^2=\Psi_1(r,\vartheta)\,dt^2-\frac{dr^2}{\Psi_2(r,\vartheta)}-r^2\Psi_3(r,\vartheta)\Big[ d\vartheta^2+\sin^2\vartheta\left(d\varphi-\Psi_4(r,\vartheta)dt\right)^2\Big]\,,
\end{equation}
where $(t,r,\vartheta,\varphi)$ refer to spherical coordinates and $\{\Psi_{i}\}_{i=1}^{4}$ are four arbitrary functions depending on $r$ and $\vartheta$.

From a mathematical point of view, the search of solutions preserving both symmetries in metric-affine geometry turns out to be a challenging task, in which highly nonlinear computations and convenient simplifications, such as the imposition of consistency constraints or hidden symmetries, play a significant role. In this sense, the consideration of a MAG model which can display both dynamical torsion and nonmetricity fields in such solutions is especially important, since the different gravitational models of MAG generally show a large number of fundamental differences that can prevent this possibility, even for static and spherically symmetric configurations~\cite{Neville:1979fk,Rauch:1981tva,Obukhov:2022khx}.

Following these lines, recently it has been shown that a rotating Kerr-Newman space-time with dynamical torsion and nonmetricity tensors holds in the decoupling limit between the orbital and the spin angular momentum~\cite{Bahamonde:2021qjk}, which raises the search of a black hole solution displaying a gravitational spin-orbit interaction beyond this limit. In this regard, an interesting route to address this problem is to explore whether or not the existing hidden symmetries of the main black hole solutions of GR can also be satisfied in the present case. In particular, the most general type D black hole solution of GR with two expanding and double principal null directions admits a nondegenerate conformal Killing-Yano tensor, which allows a complete integrability of geodesics and a separability of wave equations~\cite{Frolov:2006pe,Kubiznak:2007kh,Houri:2007xz,Oota:2007vx,Frolov:2017kze,Frolov:2017whj,Frolov:2018pys,Krtous:2018bvk,Frolov:2018ezx} (see~\cite{Podolsky:2021zwr,Podolsky:2022xxd} for new and especially convenient forms of the solution in the framework of GR, which considerably simplify the study of its physical properties). Thereby, in this section we shall consider the algebraic classification we obtained previously and show that under the slow-rotation approximation the field strength tensors of torsion of the complete solution cannot present double-aligned principal null directions preserving the type D algebraic structure of the space-time. 

In order to demonstrate this result, we can express the gravitational action of the purely torsional sector of the model of the solution as
\begin{equation}\label{ModelAction}
S = \frac{1}{64\pi}\int \dd^4x \sqrt{-\,g}
\left.\Bigl(
-\,4R-6d_{1}\tilde{R}_{\lambda\left[\rho\mu\nu\right]}\tilde{R}^{\lambda\left[\rho\mu\nu\right]}-9d_{1}\tilde{R}_{\lambda\left[\rho\mu\nu\right]}\tilde{R}^{\mu\left[\lambda\nu\rho\right]}+8\,d_{1}\tilde{R}_{\left[\mu\nu\right]}\tilde{R}^{\left[\mu\nu\right]}
\Bigr.
\right.\Bigl.\Bigr)\,,
\end{equation}
or, equivalently
\begin{equation}\label{Lagrangian2}
S = \frac{1}{64\pi}\int \dd^4x \sqrt{-\,g}
\left.\Bigl(
-\,4R-9d_1 {\nearrow\!\!\!\!\!\!\!\tilde{R}}^{(T)}_{\lambda[\rho\mu\nu]}{\nearrow\!\!\!\!\!\!\!\tilde{R}}^{(T)}{}^{\lambda[\rho\mu\nu]}+2d_1 \tilde{R}^{(T)}_{[\mu\nu]}\tilde{R}^{(T)}{}^{[\mu\nu]}-\frac{d_1}{8}\ast\tilde{R}^2
\Bigr.
\right.\Bigl.\Bigr)\,.
\end{equation}
The metric and connection field equations can then be written in terms of the antisymmetrised part of the curvature tensor and the antisymmetric Ricci tensor in the following way:
\begin{equation}\label{fieldeq3}
    G_{\rho}{}^{\sigma}=\frac{3}{2}d_1\left(3\tilde{R}_{\lambda[\rho\mu\nu]}\tilde{R}^{\lambda[\sigma\mu\nu]}-\tilde{R}_{\rho[\lambda\mu\nu]}\tilde{R}^{\sigma[\lambda\mu\nu]}\right)+\frac{9}{2}d_1\tilde{R}_{\lambda[\rho\mu\nu]}\tilde{R}^{\mu[\sigma\lambda\nu]}-4d_1\tilde{R}_{[\rho\lambda]}\tilde{R}^{[\sigma\lambda]}+8\pi\tilde{\mathcal{L}}\,\delta_{\rho}{}^{\sigma}\,,
\end{equation}
\begin{eqnarray}\label{fieldeq2}
0 &=& 2d_{1}\Bigl\{
\nabla_{\rho}\left[g^{\mu\nu}\tilde{R}^{\left[\lambda\rho\right]}-g^{\lambda\nu}\tilde{R}^{\left[\mu\rho\right]}+g^{\lambda\rho}\tilde{R}^{\left[\mu\nu\right]}-g^{\mu\rho}\tilde{R}^{\left[\lambda\nu\right]}\right]-g^{\lambda\nu}K^{\mu}\,_{\sigma\rho}\tilde{R}^{[\sigma\rho]}-g^{\mu\nu}K_{\sigma}\,^{\lambda}\,_{\rho}\tilde{R}^{[\sigma\rho]}
\Bigr.\,
\nonumber\\
\Bigl.
&&+\,K^{\nu\lambda}\,_{\rho}\tilde{R}^{[\mu\rho]}+K^{\mu}\,_{\rho}\,^{\lambda}\tilde{R}^{[\rho\nu]}-K^{\rho\lambda}\,_{\rho}\tilde{R}^{[\mu\nu]}+K^{\mu\nu}\,_{\rho}\tilde{R}^{[\lambda\rho]}+K_{\rho}\,^{\lambda\mu}\tilde{R}^{[\rho\nu]}-K^{\mu\rho}\,_{\rho}\tilde{R}^{[\lambda\nu]}
\Bigr\}\nonumber\\
&&+\,3d_{1}\Bigl\{
\nabla_{\rho}\Bigl(\tilde{R}^{\rho[\lambda\mu\nu]}+\frac{1}{2}\tilde{R}^{\lambda[\rho\mu\nu]}-\frac{1}{2}\tilde{R}^{\mu[\rho\lambda\nu]}-\tilde{R}^{\nu[\rho\lambda\mu]}\Bigr)+\,K^{\mu}\,_{\sigma\rho}\Bigl(\tilde{R}^{\rho[\lambda\sigma\nu]}+\frac{1}{2}\tilde{R}^{\lambda[\rho\sigma\nu]}-\frac{1}{2}\tilde{R}^{\sigma[\rho\lambda\nu]}-\tilde{R}^{\nu[\rho\lambda\sigma]}\Bigr)
\Bigr.\,
\nonumber\\
\Bigl.
&&-K_{\sigma}\,^{\lambda}\,_{\rho}\Bigl(\tilde{R}^{\rho[\sigma\mu\nu]}+\frac{1}{2}\tilde{R}^{\sigma[\rho\mu\nu]}-\frac{1}{2}\tilde{R}^{\mu[\rho\sigma\nu]}-\tilde{R}^{\nu[\rho\sigma\mu]}\Bigr)
\Bigr\}\,,
\end{eqnarray}
where $\tilde{\mathcal{L}}$ represents the quadratic order of the Lagrangian density. A first look at the metric field equation~\eqref{fieldeq3} clearly shows that the corrections to the Einstein equations provided by the torsion tensor of the model are traceless, which means that the corresponding stationary and axisymmetric solution must be scalar-flat, in the sense that the Riemannian Ricci scalar vanishes.

It is worthwhile to stress that the form of the antisymmetrised part of the curvature tensor that solves the metric field equation~\eqref{fieldeq3} can be straightforwardly determined by assuming the slow-rotation approximation. In this case, the metric tensor and the antisymmetrised part of the curvature tensor of the solution can be described up to the first order of the parameter $a$ by three functions $\{f_{i}\}_{i=1}^{3}$ and sixteen functions $\{a_{i}\}_{i=1}^{16}$, respectively:
\begin{align}\label{eq:metric}
    \dd s^2&=\Big(\Psi(r)+a\, f_1(r,\vartheta)\Big) dt^2-\Big(\frac{1}{\Psi(r)}+a\, \frac{f_2(r,\vartheta)}{\Psi(r)^2}\Big)d r^2- r^2 \sin ^2\vartheta \, d\varphi^2+2a\Big(1-\Psi(r)-f_3(r,\vartheta)\Big)\sin ^2\vartheta\, dt \, d\varphi\,,
\end{align}
\begin{eqnarray}\label{Ranti1}
    \tilde{R}^t{}_{[tr\vartheta]}&=&a\,a_1(r,\vartheta)\,,\quad   \tilde{R}^r{}_{[tr\vartheta]}=a\, a_2(r,\vartheta)\,,\quad  \tilde{R}^\vartheta{}_{[tr\vartheta]}=a\,a_3(r,\vartheta)\,,\quad  \tilde{R}^\varphi{}_{[tr\vartheta]}=a\,a_4(r,\vartheta)\,,\quad \\
    \tilde{R}^t{}_{[tr\varphi]}&=&a\,a_5(r,\vartheta)\,,\quad  \tilde{R}^r{}_{[tr\varphi]}=a\,a_6(r,\vartheta)\,,\quad \tilde{R}^\vartheta{}_{[tr\varphi]}=a\,a_7(r,\vartheta)\,,\quad \tilde{R}^\varphi{}_{[tr\varphi]}=a\,a_8(r,\vartheta)\,,\\ 
    \tilde{R}^t{}_{[t\vartheta\varphi]}&=&-\,\frac{1}{3}\kappa_{\rm s}  \sin\vartheta+a\,a_9(r,\vartheta)\,, \tilde{R}^r{}_{[t\vartheta\varphi]}=\frac{1}{3}\Psi(r)\kappa_{\rm s}  \sin\vartheta+a\,a_{10}(r,\vartheta)\,,\,\, \tilde{R}^\vartheta{}_{[t\vartheta\varphi]}=a\,a_{11}(r,\vartheta)\,,\,\, \tilde{R}^\varphi{}_{[t\vartheta\varphi]}=a\,a_{12}(r,\vartheta)\,,\nonumber \\ \\
    \tilde{R}^t{}_{[r\vartheta\varphi]}&=&\frac{\kappa_{\rm s} \sin\vartheta}{3\Psi(r)}+a\,a_{13}(r,\vartheta)\,,\quad \tilde{R}^r{}_{[r\vartheta\varphi]}=-\,\frac{1}{3}\kappa_{\rm s}\sin\vartheta+a\,a_{14}(r,\vartheta)\,,\quad  \tilde{R}^\vartheta{}_{[r\vartheta\varphi]}=a\,a_{15}(r,\vartheta)\,,\quad \tilde{R}^\varphi{}_{[r\vartheta\varphi]}=a\,a_{16}(r,\vartheta)\,,\nonumber \\  \label{Ranti2}
    \end{eqnarray}
where
\begin{eqnarray}
    \Psi(r)=1-\frac{2m}{r}+\frac{d_1\kappa_{\rm s} ^2}{r^2}\,.\label{Aform}
\end{eqnarray}
Then, by expanding the metric field equation~\eqref{fieldeq3} in the slow-rotation approximation, we find that the relations
\begin{align}
    a_{9}&=a_{14}\,,\quad a_{15}=\frac{a_6+a_{11}}{\Psi}-a_5\,,\quad  a_{16}=\frac{a_2+a_{12}}{\Psi}-a_1\,,\label{metricsol1}\\
    d_1\kappa_{\rm s}a_1\Psi&=\biggl\{\frac{1}{6}\Bigl[\Psi\left(2f_3-r^{2}f_{3,rr}\right)-f_{3,\vartheta\vartheta}\Bigr]+\frac{2d_1\kappa_{s}^2\left(f_3-1\right)}{3r^2}\bigg\}\sin\vartheta-\frac{1}{2}\cos\vartheta f_{3,\vartheta}+d_1\kappa_{\rm s} a_2\,,\\
   d_1\kappa_{\rm s}a_4\Psi\sin^2\vartheta&=\frac{1}{12}\bigg\{r^2\Psi'\left(f_{1,r}-f_{2,r}\right)-2\Psi\left[r^{2}f_{1,rr}+2r(f_{2,r}+f_{1,r})+2f_1\right]+\left(f_{2,\vartheta\vartheta}-f_{1,\vartheta\vartheta}\right)\nonumber\\
   &+\left(f_2-f_1\right)\left[\frac{r^2 \Psi'^2}{\Psi}-4r\Psi'-8 \Psi+4-\frac{8d_1\kappa_{s}^2}{r^2}\right]\bigg\}\sin\vartheta+\frac{1}{4}\left(f_{2,\vartheta}-f_{1,\vartheta}\right)\cos\vartheta-d_1\kappa_{\rm s}a_7\Psi\,,\\
     d_1\kappa_{\rm s}a_5\Psi^2&=\frac{1}{12}\,r\sin\vartheta\Big[2r\Psi f_{1,r\vartheta}-\left(r \Psi'+2 \Psi\right) \left(f_{1,\vartheta}-f_{2,\vartheta}\right)\Big]+d_1 \kappa_{\rm s}a_6\Psi\,,\\
    d_1\kappa_{\rm s}a_{10}&=\bigg\{\frac{1}{24}r^2\Bigl[2r\Bigl(\Psi' \left(f_1-f_2\right)+\Psi\left(f_{2,r}-f_{1,r}\right)\Bigr)-f_{1,\vartheta\vartheta}-f_{2,\vartheta\vartheta}\Bigr]+\frac{1}{3} d_1 \kappa_{s}^2 (f_1+f_2)\biggr\}\sin\vartheta\nonumber\\
   &-\,\frac{1}{24}r^2\cos\vartheta\left(f_{1,\vartheta}+f_{2,\vartheta}\right)+d_1\kappa_{\rm s}a_{13}\Psi^2\,,\\
   d_1\kappa_{\rm s}a_{14}\Psi&=\bigg\{\frac{1}{48}r^2\Bigl[6r\Psi'\left(f_2-f_1\right)+2\Psi\left(3r f_{1,r}+5r f_{2,r}+8f_2\right)+3f_{1,\vartheta\vartheta}-5f_{2,\vartheta\vartheta}\Bigr]+\frac{1}{6}d_1\kappa_{s}^2\left(f_1+f_2\right)\bigg\}\sin\vartheta\nonumber\\
   &+\frac{1}{48}r^2\cos\vartheta\left(3f_{1,\vartheta}-5f_{2,\vartheta}\right)+d_1\kappa_{\rm s}a_{13}\Psi^2\,.\label{metricsol2}
\end{align}
turn out to solve all the components of this equation, except the component $\varphi$--$\varphi$ that essentially describes the scalar-flat condition of the space-time, namely:
\begin{eqnarray}\label{f1f2}
   0&=&r\left\{2\Psi\left[r f_{1,rr}+2\left(f_{1,r}+f_{2,r}\right)\right]+r\Psi'\left(f_{2,r}-f_{1,r}\right)\right\}+2\left(f_{1,\vartheta\vartheta}-f_{2,\vartheta\vartheta}\right)+2\cot\vartheta\left(f_{1,\vartheta}-f_{2,\vartheta}\right)\nonumber\\
   &&+\left(f_{1}-f_{2}\right)\left(\frac{r^2\Psi'^2}{\Psi}-4r\Psi'-4\Psi-\frac{8d_1\kappa_{\rm s}^2}{r^2}+4\right)+4\Psi f_2\,.
\end{eqnarray}
For simplicity in the notation, we use prime to denote total derivatives of the metric function $\Psi(r)$ and comma for the partial derivatives of the additional metric functions $f_i(r,\vartheta)$ with respect to $r$ or $\vartheta$ (e.g. $f_{3,r\vartheta}=\partial^2 f_3/\partial r \partial \vartheta$).

On the other hand, the expressions for three metric functions $f_i(r,\vartheta)$ associated with a type D Riemannian Weyl tensor turn out to be further constrained. In this sense, it is worthwhile to stress that the static and spherically symmetric part of the solution trivially satisfies this condition, but this is not the case for a general stationary and axisymmetric configuration. Therefore, preserving the type D algebraic structure of the space-time from staticity and spherical symmetry to stationarity and axial symmetry constitutes an additional assumption that the complete solution may or may not satisfy.

In order to shed light on this possibility, we introduce the null vectors reproducing the slowly rotating space-time described by the line element~\eqref{eq:metric} as follows:
\begin{eqnarray}\label{nullt1}
    k^\mu&=&\bigg\{\frac{1}{\sqrt{2\Psi}}-\frac{a f_1}{2 \sqrt{2\Psi} \Psi}+\frac{a F_1}{\sqrt{2}r\Psi},\sqrt{\frac{\Psi}{2}}+\frac{a f_2}{2 \sqrt{2\Psi}}+\frac{a F_1}{\sqrt{2} r},-\,a\left(G_{1}-G_{3}\right)\sin\vartheta,a G_{4}\bigg\}\,,\\
    l^\mu&=&\bigg\{\frac{1}{\sqrt{2\Psi}}-\frac{a f_1}{2 \sqrt{2\Psi} \Psi}-\frac{a F_1}{\sqrt{2} r \Psi},-\,\sqrt{\frac{\Psi}{2}}-\frac{a f_2}{2 \sqrt{2\Psi}}+\frac{a F_1}{\sqrt{2} r},a G_{1}\sin\vartheta,a G_{2}\bigg\}\,,\\
    m^\mu&=&\bigg\{a\bigg[\frac{1}{\sqrt{2}r}+\frac{f_3-1}{\sqrt{2}r\Psi}+\frac{r\left(G_2+G_4+i G_3\right)}{2\sqrt{\Psi}}\bigg]\sin\vartheta \,,\frac{1}{2}ar\left(G_2-G_4+2iG_1-iG_3\right)\sqrt{\Psi}\sin\vartheta\,,\nonumber\\
    &&\frac{i}{\sqrt{2}r}-\frac{a F_2 \sin\vartheta}{\sqrt{2} r}\,,\frac{\csc\vartheta}{\sqrt{2} r}+\frac{ia F_2}{\sqrt{2} r}\bigg\}\\
    \bar{m}^\mu&=&\bigg\{a\bigg[\frac{1}{\sqrt{2}r}+\frac{f_3-1}{\sqrt{2}r\Psi}+\frac{r\left(G_2+G_4-i G_3\right)}{2\sqrt{\Psi}}\bigg]\sin\vartheta ,\frac{1}{2}ar\left(G_2-G_4-2iG_1+i G_3\right)\sqrt{\Psi} \sin\vartheta\,,\nonumber\\
    &&-\,\frac{i}{\sqrt{2}r}-\frac{a F_2 \sin\vartheta}{\sqrt{2}r},\frac{\csc\vartheta}{\sqrt{2}r}-\frac{ia F_2}{\sqrt{2}r}\bigg\}\,,\label{nullt4}
\end{eqnarray}
where $\{F_i\}_{i=1}^{2}$ and $\{G_i\}_{i=1}^4$ are six arbitrary real functions depending on the coordinates $r$ and $\vartheta$. Then, the condition such that the null vectors represent principal null directions of the Riemannian Weyl tensor
\begin{eqnarray}
   k_{[\sigma}W_{\lambda] \rho \mu [\nu}k_{\omega]}k^{\rho}k^{\mu}=l_{[\sigma}W_{\lambda] \rho \mu [\nu}l_{\omega]}l^{\rho}l^{\mu}=m_{[\sigma}W_{\lambda] \rho \mu [\nu}m_{\omega]}m^{\rho}m^{\mu}=\bar{m}_{[\sigma}W_{\lambda] \rho \mu [\nu}\bar{m}_{\omega]}\bar{m}^{\rho}\bar{m}^{\mu}=0\,,
   \label{cond1}
   \end{eqnarray}
is reduced to the following differential equations involving the metric functions:
\begin{equation}
    f_{1,\vartheta\vartheta}+f_{2,\vartheta\vartheta}=\cot\vartheta\left(f_{1,\vartheta}+f_{2,\vartheta}\right)\,, \quad f_{3,r\vartheta}=\frac{\Psi'}{\Psi}f_{3,\vartheta}\,,\label{f3}
\end{equation}
which allows us to find the following solution for $f_3$:
\begin{eqnarray}
    f_3(r,\vartheta)&=&f_{3a}(r)+\Psi(r)f_{3b}(\vartheta)\,,\label{formf3}
\end{eqnarray}
where $f_{3a}$ and $f_{3b}$ are two integration functions. Additionally, the condition for the Riemannian Weyl tensor to be type D reads:
   \begin{eqnarray}
    W_{\lambda \rho \mu [\nu}k_{\omega]}k^{\rho}k^{\mu}=W_{\lambda \rho \mu [\nu}l_{\omega]}l^{\rho}l^{\mu}=0\,,\label{cond2}
\end{eqnarray}
which determines the form of the functions $G_i$ as follows:
 \begin{eqnarray}
                 G_1(r,\vartheta)&=&\frac{r \Psi' \left(f_{1,\vartheta}-f_{2,\vartheta}\right)+2 \Psi \left(f_{1,\vartheta}-r f_{1,r\vartheta}+f_{2,\vartheta}\right)}{2 \sqrt{2\Psi} r \Psi \sin\vartheta\left(r^2 \Psi''-2 r \Psi'+2 \Psi-2\right)}\,,\quad G_3(r,\vartheta)=0\,,\label{G1}\\
                G_2(r,\vartheta)&=&G_4(r,\vartheta)=\frac{f_3 \left(2 r \Psi'+2-r^2 \Psi''\right)+\Psi \left(r^2 f_{3,rr}-2 r f_{3,r}+2\right)-f_{3,\vartheta\vartheta}-3 \cot \vartheta f_{3,\vartheta}+r^2 \Psi''-2 r \Psi'-2}{\sqrt{2\Psi} r^2  \left(r^2 \Psi''-2 r \Psi'+2 \Psi-2\right)}\,.\nonumber\\ \label{G2}
             \end{eqnarray}
Thus, the null vectors satisfying the conditions~\eqref{cond1} and~\eqref{cond2} read
\begin{eqnarray}
    k^\mu&=&\bigg\{\frac{1}{\sqrt{2\Psi}}-\frac{a f_1}{2 \sqrt{2\Psi}\Psi}+\frac{a F_1}{\sqrt{2} r \Psi},\sqrt{\frac{\Psi}{2}}+\frac{a f_2}{2 \sqrt{2\Psi}}+\frac{a F_1}{\sqrt{2} r},-a G_{1},a G_{2}\bigg\}\,,\\
    l^\mu&=&\bigg\{\frac{1}{\sqrt{2\Psi}}-\frac{a f_1}{2 \sqrt{2\Psi} \Psi}-\frac{a F_1}{\sqrt{2} r \Psi},-\,\sqrt{\frac{\Psi}{2}}-\frac{a f_2}{2\sqrt{2\Psi}}+\frac{a F_1}{\sqrt{2}r},a G_{1},a G_{2}\bigg\}\,,\\
    m^\mu&=&\bigg\{a\bigg[\frac{1}{\sqrt{2} r}+\frac{f_3-1}{\sqrt{2}r\Psi}+\frac{G_2 r}{\sqrt{\Psi}}\bigg]\sin\vartheta,iar G_1\sqrt{\Psi},\frac{i}{\sqrt{2}r}-\frac{a F_2\sin\vartheta}{\sqrt{2}r},\frac{\csc\vartheta}{\sqrt{2}r}+\frac{ia F_2}{\sqrt{2}r}\bigg\}\\
    \bar{m}^\mu&=&\bigg\{a\bigg[\frac{1}{\sqrt{2}r}+\frac{f_3-1}{\sqrt{2}r\Psi}+\frac{G_2 r}{\sqrt{\Psi}}\bigg]\sin\vartheta,-\,i ar G_1\sqrt{\Psi},-\,\frac{i}{\sqrt{2}r}-\frac{a F_2 \sin\vartheta}{\sqrt{2}r},\frac{\csc\vartheta}{\sqrt{2}r}-\frac{ia F_2}{\sqrt{2}r}\bigg\}\,.
\end{eqnarray}
where $\{G_i\}_{i=1}^{2}$ are given by Eqs.~\eqref{G1}-\eqref{G2}, and $\{F_i\}_{i=1}^{2}$ remain as arbitrary functions depending on $r$ and $\vartheta$.

Let us now impose that the antisymmetric part of the Ricci tensor is doubly aligned with the principal null directions $k^{\mu}$ and $l^{\mu}$ of the Riemannian Weyl tensor. Such a condition has the same form as the Expression~\eqref{typeDricci}, which was obtained in Sec.~\ref{Algebraic_Class3} for the algebraic classification of the antisymmetric Ricci tensor, namely:
\begin{equation}
    \bigl(\tilde{R}^{(T)}_{[\mu\nu]}k_{\lambda}-\tilde{R}^{(T)}_{[\mu\lambda]}k_{\nu}\bigr)k^{\mu}=\bigl(\tilde{R}^{(T)}_{[\mu\nu]}l_{\lambda}-\tilde{R}^{(T)}_{[\mu\lambda]}l_{\nu}\bigr)l^{\mu}=0\,.
\end{equation}
This condition can then be easily solved, setting the following four relations:
\begin{eqnarray}
    a_{16}=-\,a_{1}\,,\quad a_{11}=-\,a_6\,,\quad a_{12}=a_2-\frac{2\sqrt{2\Psi}}{3}\,\kappa_{\rm s}  G_2\sin\vartheta \,,\quad a_{15}=a_5+\frac{4}{3\sqrt{2\Psi}}\,\kappa_{\rm s}  G_1\sin^2\vartheta\,.\label{da}
\end{eqnarray}

As for the spin connection of the solution fulfilling stationary and axisymmetric conditions, it can be referred to the tetrad field
\begin{eqnarray}
    \vartheta^a{}_\mu =\left(
\begin{array}{cccc}
\frac{1}{2} (\Psi+1)+ \frac{a (\Psi+1) f_1}{4 \Psi} & \frac{\Psi-1}{2 \Psi}-\frac{a (\Psi-1) f_2}{4 \Psi^2} & 0 & -\frac{a (\Psi+1) \sin ^2\vartheta}{2 \Psi} (f_3+\Psi-1) \\
 \frac{1}{2} (\Psi-1)+\frac{a (\Psi-1) f_1}{4 \Psi} & \frac{\Psi+1}{2 \Psi}-\frac{a (\Psi+1) f_2}{4 \Psi^2} & 0 & -\frac{a (\Psi-1) \sin ^2\vartheta  }{2 \Psi} (f_3+\Psi-1)\\
 0 & 0 & r & 0 \\
 0 & 0 & 0 & r \sin \vartheta  \\
\end{array}
\right)\,,
\end{eqnarray}in terms of $24$ independent functions $q_i(r,\vartheta)$ as follows:
\begin{align}
    w^{\hat{0}\hat{1}}&=a\left(q_1 dt+q_2 dr+q_3 d\vartheta+q_4 d\varphi\right)\,,\\
    w^{\hat{0}\hat{2}}&=-\,\frac{1}{2}d\vartheta+a\left(q_5 dt+q_6 dr+q_7 d\vartheta+q_8 d\varphi\right)\,,\\
     w^{\hat{0}\hat{3}}&=-\,\frac{1}{2}\sin\vartheta d\varphi+a\left(q_9 dt+q_{10} dr+q_{11} d\vartheta+q_{12} d\varphi\right)\,,\\
     w^{\hat{1}\hat{2}}&=\frac{1}{2}d\vartheta+a\left(q_{13} dt+q_{14} dr+q_{15} d\vartheta+q_{16} d\varphi\right)\,,\\
     w^{\hat{1}\hat{3}}&=\frac{1}{2}\sin\vartheta d\varphi+a\left(q_{17} dt+q_{18} dr+q_{19} d\vartheta+q_{20} d\varphi\right)\,,\\
     w^{\hat{2}\hat{3}}&=-\,\frac{\kappa_{\rm s} }{r}dt+\frac{\kappa_{\rm s} }{r\Psi}dr+\cos\vartheta d\varphi+a\left(q_{21} dt+q_{22} dr+q_{23} d\vartheta+q_{24} d\varphi\right)\,.
\end{align}
Then, by replacing the condition~\eqref{da} for the antisymmetric Ricci tensor, as well as the solution ~\eqref{metricsol1}-\eqref{metricsol2} of the metric field equation into the component $\vartheta$--$\varphi$--$\varphi$ of the connection field equation~\eqref{fieldeq2}, we find:
\small{\begin{eqnarray}\label{Eq_q9}
     q_9 &=&\frac{1}{24 d_1 \kappa_{\rm s} ^2 r^2 \left(r^2 \Psi+d_1 \kappa_{\rm s} ^2-r^2\right)^2}\Big[-6 (\Psi-1) \Psi \left(3 \Psi^2 r^4+3 r^4-5 d_1 \kappa_{\rm s} ^2 r^2+2 d_1^2 \kappa_{\rm s} ^4+\left(5 d_1 r^2 \kappa_{\rm s} ^2-6 r^4\right) \Psi\right) \sin \vartheta f_{3,rrr} r^5\nonumber\\
   && -18 (\Psi-1) \left(\Psi r^2-r^2+d_1 \kappa_{\rm s} ^2\right) \left(3 \Psi r^2-3 r^2+4 d_1 \kappa_{\rm s} ^2\right) \cos \vartheta f_{3,r\vartheta} r^3-6 (\Psi-1) \Big(3 \Psi^2 r^4+3 r^4-7 d_1 \kappa_{\rm s} ^2 r^2\nonumber\\
    &&+4 d_1^2 \kappa_{\rm s} ^4+\left(7 d_1 r^2 \kappa_{\rm s} ^2-6 r^4\right) \Psi\Big) \sin \vartheta f_{3,r\vartheta\vartheta} r^3+f_3 \Big\{12 r^3 (\Psi-1) \left(3 \Psi^2 r^4+3 r^4-7 d_1 \kappa_{\rm s} ^2 r^2+4 d_1^2 \kappa_{\rm s} ^4\right.\nonumber\\
    &&-6 \left(r^4-d_1 r^2 \kappa_{\rm s} ^2\right) \Psi\left.\right) \sin \vartheta \Psi'-2 \left(\right.9 \Psi^4 r^6+\left(76 d_1 r^4 \kappa_{\rm s} ^2-15 r^6\right) \Psi^3-30 d_1 \kappa_{\rm s} ^2 \left(r^2-d_1 \kappa_{\rm s} ^2\right)^2+(3 r^6-154 d_1 \kappa_{\rm s} ^2 r^4\nonumber\\
    &&+97 d_1^2 \kappa_{\rm s} ^4 r^2) \Psi^2+3 \left(r^6+36 d_1 \kappa_{\rm s} ^2 r^4-47 d_1^2 \kappa_{\rm s} ^4 r^2+14 d_1^3 \kappa_{\rm s} ^6\right) \Psi\left.\right) \sin \vartheta\Big\}+\Big\{18 d_1 \kappa_{\rm s} ^2 (\Psi-1) \cos \vartheta \Psi' r^5\nonumber\\
    &&+3 \left(3 \left(r^2+d_1 \kappa_{\rm s} ^2\right)^2+\Psi \left(\Psi \left(9 \Psi r^2-15 r^2+34 d_1 \kappa_{\rm s} ^2\right) r^2+\left(3 r^2-d_1 \kappa_{\rm s} ^2\right) \left(r^2-13 d_1 \kappa_{\rm s} ^2\right)\right)\right) \cos \vartheta r^2\Big\}f_{3,\vartheta}\nonumber\\
    &&+\Big\{6 d_1 \kappa_{\rm s} ^2 (\Psi-1) \sin \vartheta \Psi' r^5+\left(\right.9 \Psi^3 r^6+3 \left(r^3+d_1 \kappa_{\rm s} ^2 r\right)^2+\left(34 d_1 r^4 \kappa_{\rm s} ^2-15 r^6\right) \Psi^2+\left(3 r^6-40 d_1 \kappa_{\rm s} ^2 r^4\right.\nonumber\\
    &&+13 d_1^2 \kappa_{\rm s} ^4 r^2\left.\right) \Psi\left.\right) \sin \vartheta\Big\} f_{3,\vartheta\vartheta}+\Big\{12 d_1 \kappa_{\rm s} ^2 \left(r^2-d_1 \kappa_{\rm s} ^2\right) (\Psi-1) \sin \vartheta \Psi' r^4+4 \left(\right.9 \Psi^4 r^6+\left(35 d_1 r^4 \kappa_{\rm s} ^2-27 r^6\right) \Psi^3\nonumber\\
    &&-9 d_1 \kappa_{\rm s} ^2 \left(r^2-d_1 \kappa_{\rm s} ^2\right)^2+\left(27 r^6-77 d_1 \kappa_{\rm s} ^2 r^4+29 d_1^2 \kappa_{\rm s} ^4 r^2\right) \Psi^2+\left(-9 r^6+51 d_1 \kappa_{\rm s} ^2 r^4-45 d_1^2 \kappa_{\rm s} ^4 r^2+9 d_1^3 \kappa_{\rm s} ^6\right) \Psi\left.\right) \sin \vartheta r\Big\} f_{3,r}\nonumber\\
    &&+12 d_1 \kappa_{\rm s}  \left(\Psi r^2-r^2+d_1 \kappa_{\rm s} ^2\right)^2 \Big\{q_{13,r} r^3+q_{5,r} r^3+q_{13} r^2+q_{5} r^2-2 q_{1,\vartheta} r^2-5 \kappa_{\rm s}  \sin \vartheta+\Psi \left(\right.-q_{13,r} r^3+q_{5,r} r^3-2 \kappa_{\rm s}  q_{10} r^2\nonumber\\
    &&+7 \kappa_{\rm s}  \sin \vartheta\left.\right)\Big\}+\Big\{r^4 \Psi \left(\right.-27 \Psi^3 r^4+39 r^4-90 d_1 \kappa_{\rm s} ^2 r^2+39 d_1^2 \kappa_{\rm s} ^4+\left(93 r^4-70 d_1 r^2 \kappa_{\rm s} ^2\right) \Psi^2+\left(\right.-105 r^4+160 d_1 \kappa_{\rm s} ^2 r^2\nonumber\\
    &&-31 d_1^2 \kappa_{\rm s} ^4\left.\right) \Psi\left.\right) \sin \vartheta-6 r^5 (\Psi-1) \left(3 \Psi^2 r^4+3 r^4-5 d_1 \kappa_{\rm s} ^2 r^2+2 d_1^2 \kappa_{\rm s} ^4-6 \left(r^4-d_1 r^2 \kappa_{\rm s} ^2\right) \Psi\right) \sin \vartheta \Psi'\Big\} f_{3,rr}\Big]\,,
\end{eqnarray}}\normalsize
where we have further used the components $\tilde{R}^{t}{}_{[t r\vartheta]}$, $\tilde{R}^{r}{}_{[tr\vartheta]}$, $\tilde{R}^{\varphi}{}_{[t\vartheta\varphi]}$ and $\tilde{R}^{\varphi}{}_{[r\vartheta \varphi]}$ of the antisymmetrised curvature tensor defined in~\eqref{Ranti1}-\eqref{Ranti2} to rewrite the expression in terms of $q_i$ instead of $a_i$. By doing the same substitution and using Eq.~\eqref{Eq_q9} in the component $t$--$r$--$\vartheta$ of the connection field equation~\eqref{fieldeq2}, we arrive at
\begin{eqnarray}\label{finalequation}
    0&=&192 d_1 r^2\Psi\sin\vartheta\left(\kappa_{\rm s}  m r-d_1 \kappa_{\rm s} ^3\right)^2-4 r^7\Psi^2 f_{3,rrr}\sin\vartheta\left(9 m r-8 d_1 \kappa_{\rm s} ^2\right)\left(m r-d_1 \kappa_{\rm s} ^2\right)\nonumber\\
   &&-\,4r^2\left(f_{3,\vartheta\vartheta}\sin\vartheta+3 f_{3,\vartheta}\cos\vartheta\right)\left[-20 d_1^3 \kappa_{\rm s} ^6+59 d_1^2 \kappa_{\rm s} ^4 m r+d_1 \kappa_{\rm s} ^2 m r^2 (r-58 m)+18 m^3 r^3\right]\nonumber\\
    &&-\,4r^4\Psi f_{3,rr}\sin\vartheta\left[2 d_1^3 \kappa_{\rm s} ^6+d_1^2 \kappa_{\rm s} ^4 r (18 r-23 m)+d_1 \kappa_{\rm s} ^2 m r^2 (40 m-37 r)+18 m^2 r^3 (r-m)\right]\nonumber\\
    &&+\,8r^3\Psi f_{3,r}\sin\vartheta\left[23 d_1^3 \kappa_{\rm s} ^6+d_1^2 \kappa_{\rm s} ^4 r (9 r-65 m)+d_1 \kappa_{\rm s} ^2 m r^2 (61 m-19 r)+9 m^2 r^3 (r-2 m)\right]\nonumber\\
    &&+\,8r^2\Psi f_3\sin\vartheta\left[-44 d_1^3 \kappa_{\rm s}^6+107 d_1^2 \kappa_{\rm s} ^4 m r+d_1 \kappa_{\rm s} ^2 m r^2 (r-82 m)+18 m^3 r^3\right]\,.
\end{eqnarray}
where we have further used Eq.~\eqref{f3} and replaced the form of the metric function $\Psi(r)$ as~\eqref{Aform}. From this expression, it is straightforward to note that the case where $f_3=0$ is not satisfied, which means that the previous conditions are incompatible with the field equations of the model for that case. If we consider the nontrivial form~\eqref{formf3} for the function $f_3$, it is easy find that for $f_{3a}(r)=0$ the aforementioned function is $f_{3}(r,\vartheta)=\Psi(r)\left(c_1+c_2\cos\vartheta\right)\csc^2\vartheta$ when $d_1\kappa_s=0$, where $c'_1$ and $c'_2$ are two integration constants. However, when  $d_1\kappa_s \neq 0$ and  $f_{3a}(r)=0$, the equation has no solutions. Finally, if $f_{3a}(r) \neq 0$, then it is possible to find the following nontrivial solution to Eq.~\eqref{finalequation}:
\begin{equation}
    f_{3}(r,\vartheta)=1+\left(c_1-1\right)\Psi(r)+c_{3}r^{2}+\Psi(r)\left[c_{1}\left(1+\cos^{2}\vartheta\right)+c_{2}\right]\csc^{2}\vartheta\,,
\end{equation}
where $\{c_{i}\}_{i=1}^{3}$ are three integration constants. Nevertheless, this function modifies the off-diagonal component of the metric tensor as
\begin{equation}
    g_{t\varphi}=-\,a\left[\left(2c_{1}+c_{2}\right)\Psi(r)+c_{3}r^{2}\sin^{2}\vartheta\right]\,,
\end{equation}
which clearly switches off the contribution of the Kerr angular momentum to the geometry of the space-time. In fact, it can be shown that the case with $c_1=c_3=0$ provides a simple solution of the field equations for
\begin{equation}
    q_3(r,\vartheta)=q_3(\vartheta)\,,\quad q_{24}(r,\vartheta)=\frac{c_2}{r}+q_3(\vartheta)\sin\vartheta\,,
\end{equation}
whereas the rest of axisymmetric metric and connection functions are zero, which reduces the form of the metric tensor as
\begin{align}
    \dd s^2&=\Psi(r) dt^2-\frac{d r^2}{\Psi(r)}- r^2 \sin ^2\vartheta \, d\varphi^2-2a c_2\Psi(r)\, dt \, d\varphi\,.
\end{align}

Therefore, for a circular and slowly rotating configuration, the antisymmetric part of the Ricci tensor of the complete solution cannot present double-aligned principal null directions $k^{\mu}$ and $l^{\mu}$ preserving the type D algebraic structure of the Kerr-Newman space-time. In this sense, it is important to stress that our analysis does not exclude the possibility in which the antisymmetric Ricci tensor is aligned with respect to one real null vector alone. Indeed, other solutions to the Einstein-Maxwell equations with an electromagnetic field which is only aligned to one of the principal null directions of the Weyl tensor are known in the framework of GR~\cite{Kowalczynski:1977wg}. Likewise, black hole geometries are not only restricted to algebraically special configurations, but algebraically general solutions have also been reported in the literature~\cite{Chng:2006gh,Achour:2021pla,Astorino:2023elf,Barrientos:2023tqb}. Thus, assuming that the gravitational spin-orbit interaction can be treated as a perturbative effect, in agreement with the slow-rotation approximation, the final form of the solution must present a rich algebraic structure, which at the same time represents a mathematical challenge for the resolution of the field equations in stationary and axisymmetric space-times.

\section{Conclusions}\label{sec:conclusions}

The role of algebraic classification in GR has shown to be of great importance in order to unravel different aspects of the theory and to find out relevant solutions to the Einstein's field equations, such as the Kerr solution describing the geometry of a rotating black hole~\cite{Stephani:2003tm,Griffiths:2009dfa}. Therefore, the same is expected to hold in MAG, where the torsion and nonmetricity tensors are included in the geometrical scheme.

Following these lines, in this work we have addressed the problem of algebraic classification in MAG and obtained the different algebraic types of the irreducible parts of the curvature tensor in Weyl-Cartan geometry. For this task, first we have considered the irreducible decomposition of the curvature tensor in general metric-affine geometries, which can be associated with a list of $11$ building blocks~\eqref{list_of_building_blocks} that include both Riemannian and post-Riemannian quantities. Particularisation to Weyl-Cartan geometry reduces then the number of building blocks to the list~\eqref{list_of_building_blocksWC}, whose nontrivial algebraic types are displayed in Tables~\ref{tab:Algebraictypes1},~\ref{tab:Algebraictypes2},~\ref{tab:Algebraictypes3},~\ref{tab:Algebraictypes4} and~\ref{tab:Algebraictypes5}. In summary, the tensors ${\nearrow\!\!\!\!\!\!\!\tilde{R}}^{(T)}_{\lambda[\rho\mu\nu]}$ and ${\nearrow\!\!\!\!\!\!\!\tilde{R}}_{(\mu\nu)}$ turn out to have $15$ possible algebraic types, the tensor $^{(1)}\tilde{W}_{\lambda\rho\mu\nu}$ has $6$, and the tensors $\tilde{R}^{(T)}_{[\mu\nu]}$ and $\tilde{R}^{\lambda}{}_{\lambda\mu\nu}$ have $3$. In this sense, the physical interpretation of all these algebraic types in accordance with the torsion and nonmetricity tensors is still unknown and should be clarified as new solutions to the field equations of MAG appear in the literature.

Accordingly, the results are expected to serve as a guiding principle in the search and analysis of solutions in the presence of dynamical torsion and nonmetricity, such as black holes and stars with intrinsic hypermomentum, gravitational waves and cosmological systems. As an application, we consider a MAG model which embodies a Kerr-Newman black hole in the decoupling limit between the orbital and the spin angular momentum~\cite{Bahamonde:2021qjk}, in order to demonstrate that the standard type D algebraic structure of this solution cannot be preserved beyond the decoupling limit, where the gravitational spin-orbit interaction is sufficiently important to modify the geometry of the space-time.

\bigskip
\bigskip
\noindent
\section*{Acknowledgements}

This work was supported by JSPS Postdoctoral Fellowships for Research in Japan and KAKENHI Grant-in-Aid for
Scientific Research (Grants No. JP21F21789 and No. JP22F22044). J.G.V. acknowledges the European Regional Development Fund through the Center of Excellence TK133 ``The Dark Side of the Universe".

\newpage

\appendix
\section{Building blocks of the irreducible decomposition of the curvature tensor in general metric-affine geometries}\label{sec:buildingblocks}

The expressions of the $11$ building blocks of the irreducible decomposition~\eqref{irrdec} in general metric-affine geometries, can be written in terms of the torsion and nonmetricity tensors as follows
\begin{eqnarray}
      {\nearrow\!\!\!\!\!\!\!\tilde{R}}_{(\mu\nu)}&=&{\nearrow\!\!\!\!\!\!\!{R}}_{\mu\nu}+\nabla_{\lambda}T_{(\mu\nu)}{}^{\lambda} -\nabla_{(\mu}T^{\lambda}{}_{\nu)\lambda}+\frac{1}{2}g_{\mu\nu}\nabla_{\lambda}T^{\rho\lambda}\,_{\rho}-\nabla_{\lambda}Q_{(\mu\nu)}{}^{\lambda}+\frac{1}{2}\nabla_{(\mu}Q_{\nu)}{}^{\lambda}{}_{\lambda}+\frac{1}{2}\nabla_{\lambda}Q^{\lambda}{}_{\mu\nu}\nonumber\\
    &&+\,\frac{1}{4}g_{\mu\nu}\bigl(\nabla_{\lambda}Q^{\rho}\,_{\rho}\,^{\lambda}-\nabla_{\lambda}Q^{\lambda}\,_{\rho}\,^{\rho}\bigr)+ \frac{1}{2}T^{\rho}{}_{\lambda(\mu}T_{\nu)}{}^{\lambda}{}_{\rho}+T^{\rho\lambda}{}_{\rho}T_{(\mu\nu)\lambda}+\frac{1}{4}T_{\mu\lambda\rho} T_{\nu}{}^{\lambda\rho}\nonumber\\
    &&+\,\frac{1}{4}g_{\mu\nu}\Bigl(T^{\lambda}\,_{\lambda\sigma}T^{\rho}\,_{\rho}\,^{\sigma}-\frac{1}{2}T_{\lambda\rho\sigma}T^{\rho\lambda\sigma}-\frac{1}{4}T_{\lambda\rho\sigma}T^{\lambda\rho\sigma}\Bigr)+Q_{\lambda\mu\rho}Q^{[\lambda}{}_{\nu}{}^{\rho]}+\frac{1}{2}Q^{\lambda \rho}{}_{\rho}Q_{(\mu\nu)\lambda}\nonumber\\
    &&-\,\frac{1}{4}\bigl(Q_{\lambda\mu\nu}Q^{\lambda\rho}{}_{\rho}+Q_{\mu\lambda\rho}Q_{\nu}{}^{\lambda\rho}\bigr)+\frac{1}{16}g_{\mu\nu}\bigl(2Q_{\lambda\rho\sigma}Q^{\rho\lambda\sigma}+Q^{\rho\lambda}\,_{\lambda}Q_{\rho}\,^{\sigma}\,_{\sigma}-Q_{\lambda\rho\sigma}Q^{\lambda\rho\sigma}-2Q^{\sigma\lambda}\,_{\lambda}Q^{\rho}\,_{\rho\sigma}\bigr)\nonumber\\
    &&+\,\frac{1}{2}\bigl(T^{\lambda}{}_{(\mu}{}^{\rho}Q_{\nu)\lambda\rho}-Q_{\lambda\rho(\mu}T^{\lambda}{}_{\nu)}{}^{\rho}-2Q_{(\mu\nu)\lambda}T^{\rho\lambda}{}_{\rho}+Q_{\lambda\rho(\mu}T^{\rho}{}_{\nu)}{}^{\lambda}-2Q_{\lambda\rho(\mu}T_{\nu)}{}^{\lambda\rho}-Q_{\lambda}{}^{\rho}{}_{\rho}T_{(\mu\nu)}{}^{\lambda}+Q_{\lambda\mu\nu}T^{\rho\lambda}{}_{\rho}\bigr)\nonumber\\
    &&+\,\frac{1}{4}g_{\mu\nu}\bigl(T^{\lambda}\,_{\lambda\rho}Q^{\rho\sigma}\,_{\sigma}-T_{\lambda\rho\sigma}Q^{\sigma\lambda\rho}-T^{\lambda}\,_{\lambda\rho}Q^{\sigma\rho}\,_{\sigma}\bigr)\,,\label{BB1}\\
    \tilde{R}^{(T)}_{[\mu\nu]}&=&\tilde{\nabla}_{[\mu}T^{\lambda}\,_{\nu]\lambda}+\frac{1}{2}\tilde{\nabla}_{\lambda}T^{\lambda}\,_{\mu\nu}-\frac{1}{2}T^{\lambda}\,_{\rho\lambda}T^{\rho}\,_{\mu\nu}\,, \quad \tilde{R}^{\lambda}\,_{\lambda\mu\nu}=\nabla_{[\nu}Q_{\mu]\lambda}{}^{\lambda}\,,\label{BB2}\\
    {\nearrow\!\!\!\!\!\!\!\hat{R}}^{(Q)}_{(\mu\nu)}&=&\tilde\nabla_{\lambda}{\nearrow\!\!\!\!\!\!\!Q}_{(\mu \nu )}{}^{\lambda}-\tilde\nabla_{(\mu}{\nearrow\!\!\!\!\!\!\!Q}^\lambda{}_{\nu )\lambda}+{\nearrow\!\!\!\!\!\!\!Q}^{\lambda\rho}\,_{\lambda}{\nearrow\!\!\!\!\!\!\!Q}_{(\mu\nu)\rho}-{\nearrow\!\!\!\!\!\!\!Q}_{\lambda\rho(\mu}{\nearrow\!\!\!\!\!\!\!Q}_{\nu)}{}^{\lambda\rho}-T_{\lambda\rho(\mu}{\nearrow\!\!\!\!\!\!\!Q}^{\lambda\rho}{}_{\nu)}\,,\label{BB3}\\
     \hat{R}^{(Q)}_{[\mu\nu]}&=&\tilde{\nabla}_{[\mu}{\nearrow\!\!\!\!\!\!\!Q}^{\lambda}{}_{\nu]\lambda}-\tilde{\nabla}_{\lambda}{\nearrow\!\!\!\!\!\!\!Q}_{[\mu\nu]}{}^{\lambda}-\frac{1}{2}\tilde{\nabla}_{[\mu}{\nearrow\!\!\!\!\!\!\!Q}_{\nu]\lambda}{}^{\lambda}+{\nearrow\!\!\!\!\!\!\!Q}_{[\nu\mu]\lambda}{\nearrow\!\!\!\!\!\!\!Q}_{\rho}{}^{\lambda\rho}-{\nearrow\!\!\!\!\!\!\!Q}_{\rho\lambda [\mu}{\nearrow\!\!\!\!\!\!\!Q}_{\nu] }{}^{\rho\lambda} + {\nearrow\!\!\!\!\!\!\!Q}_{\lambda\rho[\mu}T^{\lambda}{}_{\nu]}{}^{\rho}+ \frac{1}{4}{\nearrow\!\!\!\!\!\!\!Q}^{\lambda\rho}{}_{\rho} T_{\lambda\mu\nu} \,,\label{BB4}\\
    {\nearrow\!\!\!\!\!\!\!\tilde{R}}^{(T)}_{\lambda[\rho\mu\nu]}&=& \frac{1}{2} g_{\lambda[\rho|}\tilde{\nabla}_{\sigma}T^{\sigma}{}_{|\mu\nu]}+g_{\lambda[\rho}\tilde{\nabla}_{\mu}T^{\sigma}{}_{\nu]\sigma}-g_{\lambda\sigma}\tilde{\nabla}_{[\rho}T^{\sigma}{}_{\mu\nu]}+\frac{1}{24}\varepsilon_{\lambda\rho\mu\nu}\varepsilon_{\sigma}{}^{\alpha\beta\gamma}\bigl(\tilde{\nabla}_{\gamma}T^{\sigma}{}_{\beta\alpha}+T_{\beta\omega\gamma}T^{\omega\sigma}{}_{\alpha}\bigr)\nonumber\\
    &&+\,T_{\lambda\sigma[\rho}T^{\sigma}{}_{\mu\nu]}-\frac{1}{2}g_{\lambda[\rho}T^{\sigma}{}_{\mu\nu]} T^{\omega}{}_{\sigma\omega}\,,\label{BB5}\\
    {\nearrow\!\!\!\!\!\!\!\tilde{R}}^{(Q)}_{\lambda[\rho\mu\nu]} &=&\frac{3}{2}\Big( g_{\lambda[\rho|}\tilde{\nabla}_{\sigma}{\nearrow\!\!\!\!\!\!\!Q}_{|\mu\nu]}{}^{\sigma}-g_{\lambda[\rho}\tilde{\nabla}_{\mu}{\nearrow\!\!\!\!\!\!\!Q}^{\sigma}{}_{\nu]\sigma}-2\tilde{\nabla}_{[\rho}{\nearrow\!\!\!\!\!\!\!Q}_{\mu\nu]\lambda}+g_{\lambda[\rho}{\nearrow\!\!\!\!\!\!\!Q}_{\mu\nu]\sigma}{\nearrow\!\!\!\!\!\!\!Q}_{\omega}{}^{\sigma\omega}+g_{\lambda[\rho}{\nearrow\!\!\!\!\!\!\!Q}^{\sigma }{}_{\mu}{}^{\omega}{\nearrow\!\!\!\!\!\!\!Q}_{\nu]\sigma\omega}+{\nearrow\!\!\!\!\!\!\!Q}_{\sigma\lambda[\rho}T^{\sigma}{}_{\mu\nu]}\nonumber\\
    &&+\,g_{\lambda[\rho|}{\nearrow\!\!\!\!\!\!\!Q}_{\sigma|\mu|}{}^{\omega}T^{\sigma}{}_{\omega|\nu]}+\frac{1}{2}Q_{[\rho|\sigma}{}^{\sigma}{\nearrow\!\!\!\!\!\!\!Q}_{|\mu\nu]\lambda}\Big)\,,\label{BB6}\\  {}^{(1)}\tilde{W}_{\lambda\rho\mu\nu}&=&\tilde{R}_{[\lambda\rho]\mu\nu}-\frac{3}{4}\Big({\nearrow\!\!\!\!\!\!\!\tilde{R}}^{(T)}_{\lambda[\rho\mu\nu]}+{\nearrow\!\!\!\!\!\!\!\tilde{R}}^{(T)}_{\nu[\lambda\rho\mu]}-{\nearrow\!\!\!\!\!\!\!\tilde{R}}^{(T)}_{\rho[\lambda\mu\nu]}-{\nearrow\!\!\!\!\!\!\!\tilde{R}}^{(T)}_{\mu[\lambda\rho\nu]}\Bigr)-\frac{1}{2}\Bigl({\nearrow\!\!\!\!\!\!\!\tilde{R}}^{(Q)}_{\mu[\lambda\rho\nu]}-{\nearrow\!\!\!\!\!\!\!\tilde{R}}^{(Q)}_{\nu[\lambda\rho\mu]}\Big)+\,\frac{1}{24}\left.\ast\tilde{R}\right. \varepsilon_{\lambda\rho\mu\nu}\nonumber\\
&&-\,\frac{1}{4}\Big[g_{\lambda\mu}\left(2{\nearrow\!\!\!\!\!\!\!\tilde{R}}_{(\rho\nu)}+{\nearrow\!\!\!\!\!\!\!\hat{R}}^{(Q)}_{(\rho \nu) }\right)+g_{\rho\nu}\left(2{\nearrow\!\!\!\!\!\!\!\tilde{R}}_{(\lambda\mu)} +{\nearrow\!\!\!\!\!\!\!\hat{R}}^{(Q)}_{(\lambda\mu)}\right)-g_{\lambda\nu}\left(2 {\nearrow\!\!\!\!\!\!\!\tilde{R}}_{(\rho\mu)}+{\nearrow\!\!\!\!\!\!\!\hat{R}}^{(Q)}_{(\rho\mu)}\right)-g_{\rho\mu}\left(2{\nearrow\!\!\!\!\!\!\!\tilde{R}}_{(\lambda\nu)}+{\nearrow\!\!\!\!\!\!\!\hat{R}}^{(Q)}_{(\lambda\nu)}\right)\Big]\nonumber\\
&&-\,\frac{1}{4}\Big[g_{\lambda\mu}\Bigl(2\tilde{R}^{(T)}_{[\rho\nu]}+\hat{R}^{(Q)}_{[\rho\nu]}\Bigr)+g_{\rho\nu}\Bigl(2\tilde{R}^{(T)}_{[\lambda\mu]}+\hat{R}^{(Q)}_{[\lambda\mu]}\Bigr)-g_{\lambda\nu}\Bigl(2\tilde{R}^{(T)}_{[\rho\mu]}+\hat{R}^{(Q)}_{[\rho\mu]}\Bigr)-g_{\rho\mu}\Bigl(2\tilde{R}^{(T)}_{[\lambda\nu]}+\hat{R}^{(Q)}_{[\lambda\nu]}\Bigr)\nonumber\\
 &&+\,\tilde{R}^{\sigma}{}_{\sigma\lambda[\mu}g_{\nu]\rho}-\tilde{R}^{\sigma}{}_{\sigma\rho[\mu}g_{\nu]\lambda}\Big]-\frac{1}{6}\tilde{R}\,g_{\lambda[\mu}g_{\nu]\rho}\,,\label{BB7}\\     
   {}^{(1)}\tilde{Z}_{\lambda\rho\mu\nu} &=&\tilde{R}_{(\lambda\rho)\mu\nu}-\frac{1}{4}\Big({\nearrow\!\!\!\!\!\!\!\tilde{R}}^{(Q)}_{\lambda[\rho\mu\nu]}+{\nearrow\!\!\!\!\!\!\!\tilde{R}}^{(Q)}_{\rho[\lambda\mu\nu]}\Big)-\frac{1}{6}\Big(g_{\lambda\nu}\hat{R}^{(Q)}_{[\rho\mu]}+g_{\rho\nu}\hat{R}^{(Q)}_{[\lambda\mu]}-g_{\lambda\mu}\hat{R}^{(Q)}_{[\rho\nu]}-g_{\rho\mu}\hat{R}^{(Q)}_{[\lambda\nu]}+g_{\lambda\rho}\hat{R}^{(Q)}_{[\mu\nu]}\Big)\nonumber\\
 &&-\,\frac{1}{4}g_{\lambda\rho}\tilde{R}^{\sigma}{}_{\sigma\mu\nu}-\frac{1}{8}\Big(g_{\lambda\nu}{\nearrow\!\!\!\!\!\!\!\hat{R}}^{(Q)}_{(\rho\mu)}+g_{\rho\nu} {\nearrow\!\!\!\!\!\!\!\hat{R}}^{(Q)}_{(\lambda\mu)}-g_{\lambda\mu} {\nearrow\!\!\!\!\!\!\!\hat{R}}^{(Q)}_{(\rho\nu)}-g_{\rho\mu} {\nearrow\!\!\!\!\!\!\!\hat{R}}^{(Q)}_{(\lambda\nu)}\Big)\,,\label{BB8}\\
    \tilde{R}&=& R-2\nabla_{\mu}T^{\nu \mu}\,_{\nu}+\nabla_{\mu}Q^{\mu}\,_{\nu}\,^{\nu}-\nabla_{\mu}Q^{\nu}\,_{\nu}\,^{\mu}+\frac{1}{4}T_{\lambda \mu \nu}T^{\lambda \mu \nu}+\frac{1}{2}T_{\lambda \mu \nu}T^{\mu \lambda \nu}-T^{\lambda}\,_{\lambda\nu}T^{\mu}\,_{\mu}\,^{\nu}+T_{\lambda\mu\nu}Q^{\nu\lambda\mu}\nonumber\\
    &&+\,T^{\lambda}\,_{\lambda\nu}Q^{\mu\nu}\,_{\mu}-T^{\lambda}\,_{\lambda\nu}Q^{\nu\mu}\,_{\mu}+\frac{1}{4}Q_{\lambda\mu\nu}Q^{\lambda\mu\nu}-\frac{1}{2}Q_{\lambda\mu\nu}Q^{\mu\lambda\nu}+\frac{1}{2}Q^{\nu\lambda}\,_{\lambda}Q^{\mu}\,_{\mu\nu}-\frac{1}{4}Q^{\nu\lambda}\,_{\lambda}Q_{\nu}\,^{\mu}\,_{\mu}\,,\label{BB9}\\
\ast\tilde{R}&=&\varepsilon^{\lambda\rho\mu\nu}\Bigl(\nabla_{\lambda}T_{\rho\mu\nu}+\frac{1}{2}T^{\sigma}{}_{\lambda\rho}T_{\sigma\mu\nu}-Q_{\lambda\sigma\rho}T^{\sigma}{}_{\mu\nu}\Bigr)\,.\label{BB10}
\end{eqnarray}

It is straightforward to note that the building blocks~\eqref{BB1},~\eqref{BB7} and~\eqref{BB9} include both Riemannian and post-Riemannian contributions, whereas the rest of building blocks are purely post-Riemannian quantities. In particular, the Holst pseudoscalar and the building blocks denoted by a superscript $T$ are nontrivial only in the presence of torsion, the homothetic component of the curvature tensor is determined by the trace of nonmetricity, and the rest of building blocks denoted by a superscript $Q$, together with the building block~\eqref{BB8}, vanish in the absence of the traceless part of the nonmetricity tensor.

\bibliographystyle{utphys}
\bibliography{references}

\end{document}